\documentclass[aip,reprint]{revtex4-1}

\usepackage{graphicx}

\usepackage{xcolor}

\newcommand{\red}{\color{black}}

\draft 

\begin{document}


\title{Analyser-free, intensity-based wide-field magneto-optical microscopy} 



\author{R.~Sch\"{a}fer}
\affiliation{Leibniz Institute for Solid State and Materials Research (IFW) Dresden, Helmholtzstrasse 20, D-01069 Dresden, Germany}
\affiliation{Institute for Materials Science, Technische Universit{\"a}t Dresden, D-01062 Dresden, Germany}
\author{Peter M. Oppeneer}
\affiliation{Department of Physics and Astronomy, Uppsala University, Box 516, SE-75120 Uppsala, Sweden}
\author{Alexey Ognev}
\affiliation{School of Natural Sciences, Far Eastern Federal University, Vladivostok, Russia}
\author{Alexander Samardak}
\affiliation{School of Natural Sciences, Far Eastern Federal University, Vladivostok 690950, Russia}
\affiliation{National Research South Ural State University, Chelyabinsk, 454080, Russia}
\author{Ivan V. Soldatov}
\affiliation{Leibniz Institute for Solid State and Materials Research (IFW) Dresden, Helmholtzstrasse 20, D-01069 Dresden, Germany}
\affiliation{{Now at Institut f\"{u}r Materialwissenschaft, Technische Universit\"{a}t Darmstadt, D-64287 Darmstadt, Germany}}


\date{\today}

\begin{abstract}

In conventional Kerr- and Faraday microscopy the sample is illuminated with plane-polarised light and a magnetic domain contrast is generated by an analyser making use of the Kerr- or Faraday rotation. In this paper we demonstrate possibilities of \textit{analyser-free} magneto-optical microscopy based on magnetisation-dependent intensity modulations of the light: (i) The transverse Kerr effect can be applied for in-plane magnetised material, demonstrated for an FeSi sheet. 
(ii) Illuminating the same sample with circularly polarised light leads to a domain contrast with a different symmetry as the conventional Kerr contrast. 
(iii) Circular polarisation can also be used for perpendicularly magnetised material, demonstrated for a garnet film and an ultrathin CoFeB film. 
(iv) Plane-polarised light at a specific angle can be employed for both, in-plane and perpendicular media. 
(v) Perpendicular light incidence leads to a domain contrast on in-plane materials that is quadratic in the magnetisation and to a domain boundary contrast. 
(vi) Domain contrast can even be obtained without polariser.
{\red In cases (ii) and (iii), the contrast is generated by MCD (Magnetic Circular Dichroism), while MLD (Magnetic Linear Dichroism) is responsible for the contrast in case (v).
The domain boundary contrast is due to the magneto-optical gradient effect in metallic samples. A domain boundary contrast can also arise due to interference of phase-shifted magneto-optical amplitudes.}
An explanation of these contrast phenomena is provided in terms of Maxwell-Fresnel theory. 

\end{abstract}


\maketitle 

\section{Introduction}
\label{Introduction}

Numerous applications of the magneto-optical (MO) Kerr effect are well established, including optical magnetometry known as MOKE (Magneto-Optical Kerr Effect) magnetometry \cite{Bader1991} as well as magnetic domain imaging in wide-field 
\footnote{``Wide-field" optical microscopy is the conventional microscopy technique, in which the whole area of interest is exposed to light and then simultaneously viewed either by eye or a camera. This is different to optical scanning microscopy, a sequential imaging technique in which a spot of light is scanned relative to the specimen in a raster-like way, building up an image point by point.}
polarisation microscopes \cite{Hubert1998, Schaefer2007, McCord2015} and by laser-scanning microscopy \cite{Egelkamp1990}. All these applications are based on linearly (or plane-) polarised, incident light that --- after interaction with the magnetisation of the specimen --- is transformed into magnetically modulated light on reflection. The reflected light wave may actually be seen as a superposition of a regularly reflected amplitude $\vec{N}$ (being polarised along the same plane as the incident light, as it would occur in case of a non-magnetic material) and a magnetisation-dependent Kerr amplitude, $\vec{K}$. The Kerr amplitude is generated by the gyroelectric interaction between the magnetisation vector $\vec{m}$ and the electrical field vector $\vec{E}_\mathrm{in}$ of the incident light wave, i.e. a non-vanishing cross product $\vec{m}\times\vec{E}_\mathrm{in}$ is required \cite{Hubert1998} 
{leading to MO effects that are linear in the magnetisation $m$}. This implies constraints for the {planes of light incidence and polarisation} relative to the $\vec{m}$-vector. Three basic Kerr modes are distinguished by convention (Fig.\,\ref{Fig-Kerreffects}):
\begin{itemize}
\item 
\textit{Polar Kerr effect}. Here the magnetisation points along the surface normal and the effect is strongest (i.e.\ $\vec{m}\times\vec{E}_\mathrm{in}$ is maximal) at perpendicular incidence. The Kerr amplitude is perpendicular to $\vec{N}$, leading to a Kerr rotation by superposition with $\vec{K}$ that is the same for any polarisation direction of the incident light. 
\item 
For the \textit{Longitudinal Kerr effect}, oblique incidence of light is required to generate a (detectable) Kerr rotation. Here the $\vec{m}$-vector lies parallel to the surface and along the plane of incidence. A Kerr rotation is obtained for both, light that is polarised parallel and perpendicular to the plane of incidence, called $p$- and $s$-polarised light.
\item 
\textit{Transverse Kerr effect}. Here the magnetisation is in-plane again, but perpendicular to the plane of incidence. A Kerr amplitude is generated only for $p$-polarised light (for $s$-polarisation $\vec{m}$ would be parallel to $\vec{E}_\mathrm{in}$), but its polarisation direction is the same as that of the regularly reflected beam. 

\begin{figure}
\center
\includegraphics[width=1\linewidth,clip=true]{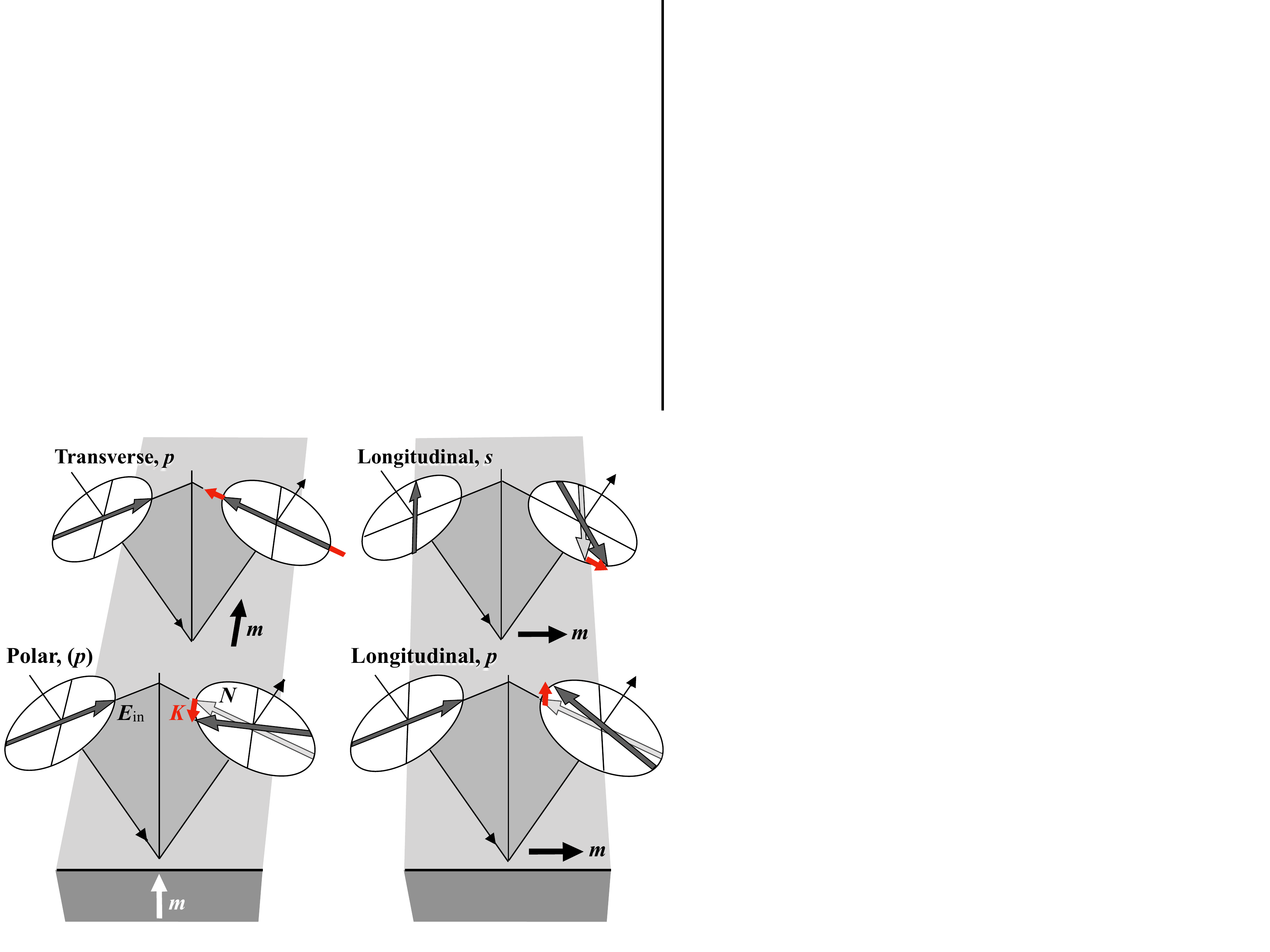}
\caption{Schematics of the basic Kerr effect modes. Shown is the magnetisation vector ($\vec{m}$), the electrical vector of the incident, plane-polarised light wave ($\vec{E}_\mathrm{in}$), the regularly reflected field amplitude ($\vec{N}$), and the Kerr amplitude ($\vec{K}$). Inversion of the magnetisation direction would lead to an inversion of the Kerr amplitude vectors.}
\label{Fig-Kerreffects}
\end{figure}

\vspace{0.05in}
\end{itemize}

The same effects are also found in transmission geometry for transparent magnetic specimens, called \textit{Faraday effects} --- with one exception: in the transverse configuration, no (linear-in-$\vec{m}$-) MO effect is possible in transmission because the cross-product is either zero or points along the propagation direction thus making a MO Faraday amplitude, $\vec{F}$, impossible. 
The Faraday signal is normally much stronger than the corresponding Kerr signal as the light interacts with the magnetisation across the whole specimen thickness and not just within some penetration depth as for the Kerr effects. Faraday microscopy is also possible in reflection geometry if the transparent sample is placed or deposited on a mirror.

In the case of the rotational effects, a detectable magneto-optical signal or domain contrast is finally obtained by using an analyser in the reflection (or transmission) path that blocks the emerging amplitude differently for differently magnetised domains, resulting in a magnetisation-dependent intensity modulation of the reflected (or transmitted) light. If there is a phase shift between $\vec{N}$ and $\vec{K}$ (or $\vec{F}$), the rotated light will also be elliptically polarised, requiring a phase-shifter (called compensator) for contrast optimisation \cite{Kuch2015}.

The Kerr- and Faraday effects are characterised by the fact that they are all, to lowest order, linear in the local magnetisation ${m}$. 
There are two other effects \cite{Kuch2015} that can lead to MO contrasts by using plane-polarised light in an optical polarisation microscope: 
The \textit{Voigt effect} depends quadratically on the magnetisation and consequently generates a contrast between domains that are magnetised in-plane and along different \textit{axes} rather than directions. As the light, emerging from the specimen, is predominantly elliptically polarised in case of the Voigt effect, the use of a compensator is mandatory in order to generate a detectable rotation. 
The MO \textit{gradient effect} finally is a birefringence effect, which depends linearly on gradients in the magnetisation vector field.

Apart from the well-known Kerr and Faraday \textit{rotation}, MO effects can also manifest themselves as a magnetically induced modification of the light's \textit{amplitude}, either in transmission or reflection geometry \cite{Oppeneer2001}. One such example is the above mention transverse Kerr effect. {As the transverse Kerr amplitude is polarised along the same plane as the regularly reflected beam (compare Fig.\,\ref{Fig-Kerreffects}), the transverse Kerr effect causes an amplitude variation of the light rather than a rotation.}
Another prominent example is the Magnetic Circular Dichroism (MCD) effect, which is customarily observed in transmission through a magnetic sample (see e.g.\ Ref.\,[\onlinecite{Mason2007}]). 
In the X-ray regime, magnetic imaging using the X-ray Magnetic Circular Dichroism (XMCD) has been successfully applied in a transmission soft X-ray microscope \cite{Fischer1996} and in combination with photoexcitation electron emission microscopy (PEEM) \cite{Stohr1998,Schneider2002}, which both permit visualisation of sub-micrometer sized ferromagnetic domains.  
The X-ray Magnetic  Linear Dichroism (XMLD) is a MO effect related to the Voigt effect that has been used to image antiferromagnetic domains in combination with PEEM \cite{Nolting2000,Scholl2000}.
These magnetic dichroic effects can be observed as well  
in a photon-in--photon-out reflection experiment when using circularly or linearly polarised X-rays (see e.g.\ Refs.\,[\onlinecite{Mertins2002}] and [\onlinecite{Valencia2010}]). The appearance of these MO effects as an amplitude variation is thus distinct from those characterised by a MO rotation, but all MO effects can be described within the common Maxwell-Fresnel framework \cite{Hecht2002, Oppeneer2001}. This, therefore, raises the question, if also the amplitude-modulation effects can be used for magnetic imaging at visible frequencies and which method is optimally suited?

Thus far, \emph{rotation-based} Kerr-, Faraday- and Voigt microscopy has been {(almost)} exclusively employed for domain imaging in the visible frequency regime \cite{Hubert1998,Schaefer2007,McCord2015}.
A recent investigation \cite{Kim2020} on ultrathin magnetic films showed that anti-reflection coatings could significantly enhance the MO rotation and further found that a MO contrast could even be detected by illuminating the sample with left- and right-circularly polarised (LCP and RCP) light rather than by plane-polarised light, thus {indeed} making use of the MCD effect. The latter contrast was enhanced by the extreme anti-reflection coating used in that work, but it indicates the possibility of magnetic domain imaging without the need for an analyser, which would not have any effect on a circularly polarised wave anyway.
{In another recent article \cite{Jin2020} the MCD effect was employed as well to image domains on an ultrathin, two-dimensional CrBr$_3$ magnetic film at low temperature. There, however, the ultrathin film was illuminated with plane-polarised light and the MCD signal was extracted after reflection by using a quarter-wave plate and an analyser. Again, the Kerr signal was significantly enhanced by interference effects with an underlayer resembling the mentioned antireflection effect. 
Interestingly, the magneto-optics of CrBr$_3$ material (on bulk crystals, though) was already investigated back in the 1960ies \cite{Dillon1966, Dillon1968}. In 1975, Kuhlow and Lambeck \cite{Kuhlow1975} published domain images of a 12\,$\mu$m thick CrBr$_3$ single crystal, imaged in blue, \emph{circularly polarised} light by Faraday microscopy at 15\,K. To our best knowledge, this is the first wide-field microscopic observation of magnetic domains making use of the MCD effect in the absence of an analyser.}
Noteworthy is also a further recent article \cite{Guilett2021} in which an alternative microscopy technique was introduced. There the MCD effect was combined with the Seebeck effect in a ferromagnet-semiconductor bilayer thus detecting the MCD signal electrically in a transmission geometry.

From an application-oriented perspective, \emph{establishing} analyser-free magnetic microscopy could be advantageous as it reduces the need for the extra optical element, provided that the magnetic modulation of the amplitude is large enough. Since these various MO effects can all be described within the Maxwell-Fresnel framework, it can be anticipated that those based on intensity modulation can indeed be employed for magnetic microscopy, given an appropriate measurement geometry, even without anti-reflection coating.

In this article we review several possibilities for analyser-free MO wide-field microscopy making use of intensity-modulated light. 
To provide the basics, we will start with the theory of the MO effects in Sect.\,\ref{Theory}, followed by some experimental details (Sect.\,\ref{Experimental}) and a section that briefly recaps conventional, analyser-based Kerr microscopy for comparison (Sect.\,\ref{Conventional Longitudinal Kerr Microscopy}).
In Sect.\,\ref{Transverse Microscopy} we then show that the amplitude modulation of  linearly polarised light due to the transverse Kerr effect can well be employed for magnetic domain imaging. 
In Sect.\,\ref{Oppeneer Effect} we report on the identification of one further MO intensity modulation configuration that is suitable for direct MO imaging when employing plane-polarised light for illumination.
In Sects.\,\ref{Longitudinal and transverse MCD-based Kerr microscopy} and \ref{Polar MCD-based Faraday microscopy} it will be demonstrated that magnetic microscopy is also possible by illuminating {\red the} magnetic specimen with circularly polarised light, i.e.\ by making direct use of the MCD effect similar to XMCD 
\cite{Stohr1998,vonKorff2014,Kuch2015,Zayko2020}. 
This is not only feasible for perpendicularly magnetised films with sophisticated antireflection coatings as mentioned, but it works for both, in-plane and perpendicular media even in the absence of antireflection coatings.
In Sect.\,\ref{Voigt Effect} we demonstrate that also the magnetic linear dichroism (MLD) effect does lead to a domain contrast, which is superimposed by a domain boundary contrast due to the magneto-optical gradient effect.
To check for eventualities, we have examined the existence of contrasts in a different microscope type with separated illumination and reflection paths (Sect.\,\ref{Overview Microscopy}).
{At the end of the paper (Sect.\,\ref{Further aspects}) we will finally mention three further aspects than might be worth to be examined in future work. This includes 
(i) the influence of the light colour on various Kerr contrasts, (ii) the finding that a domain contrast can even be seen in the absence of analyser \textit{and} polariser, and (iii) diffraction effects at domain boundaries. In this context we will also refer to largely forgotten, 60 years old research that has already pointed out the possibility of magneto-optical imaging without the need of polariser and analyser, but, however, based on dark-field optical microscopy.
\footnote{{In dark-field microscopy the sample is illuminated with light that will not be collected by the objective lens so that only scattered light contributes to the image formation. This produces an almost black background image with bright objects superimposed.}}}


\section{Theory} 
\label{Theory}

The here-studied {Kerr- and Faraday} effects are characterised by the fact that they are all, to lowest order, linear in the local magnetisation ${m}$ and can all be described on the same footing within the macroscopic Maxwell-Fresnell theory \cite{Yang1993,Hecht2002,Oppeneer2001}.
In this formulation the complex Voigt constant $Q_\mathrm{V}$ ($\sim m$) captures the MO interaction strength of the Kerr and Faraday effects.

\begin{figure}
\center
\includegraphics[width=1\linewidth,clip=true]{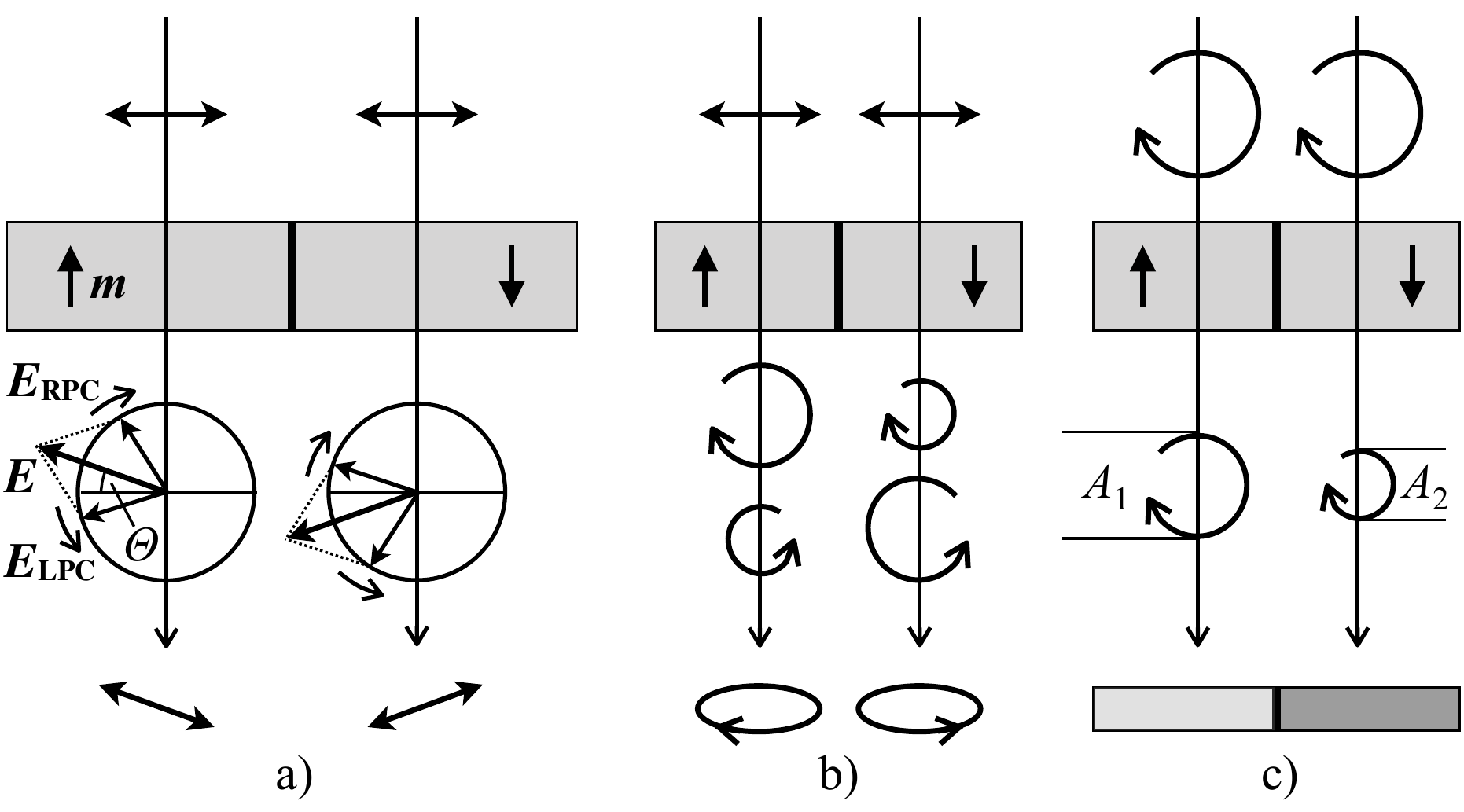}
\caption{(\textbf{a}) Schematic representation of magnetic circular birefringence or Faraday effect: plane-polarised light is resolved into left- and right-handed circular partial waves that are phase-shifted, resulting in rotated plane waves in dependence of the magnetisation direction. (\textbf{b}) 
Different damping of the two partial wave amplitudes, which is inverted by inverting the magnetisation direction, results in elliptical waves with opposite handedness (Faraday ellipticity).
Note that the effects of (a) and (b) generally occur together so that the emerging light is usually rotated \textit{and} more or less elliptically polarised. 
(\textbf{c}) MCD for the case of right-handed circular illumination. Differential, magnetisation-dependent damping of the circular wave leads to absorption domain contrast. The effects are shown for transmission (Faraday) geometry, but they similarly apply to reflection (Kerr geometry).}
\label{Fig-Sketch}
\end{figure}

To start with the MO \textit{rotation} effects, these can be understood by applying a concept that was originally proposed by Fresnel to describe optical activity \cite{Hecht2002}: A plane-polarised light wave can be represented by the superposition of  LCP and RCP partial waves with the same amplitudes. By solving the {Fresnel} wave equation for ferro- or ferrimagnetic magnetic materials, only two such circularly polarised eigenmodes with opposite rotation sense can propagate  \textit{along} the magnetisation vector within the material. 
A plane-polarised wave, entering the material, is thus resolved into these two eigenmodes (Fig.\,\ref{Fig-Sketch}).
Each mode experiences its own (complex) index of refraction, $n_\mathrm{RCP}$ or $n_\mathrm{LCP}$, during propagation. 
The difference of the indices is $n_\mathrm{RCP} - n_\mathrm{LCP} \approx - \bar{n}Q_\mathrm{V}$, with $\bar{n}$ the average index of refraction.
The two partial waves will hence advance with different velocities and amplitudes in the material.
The former result in a phase shift, which leads to a rotated plane wave by superposition of the two partial waves, the rotation sense of which depends on the magnetisation direction [Fig.\,\ref{Fig-Sketch}\,(a)]. The Faraday rotation $\theta_{\mathrm{F}}$ is thus proportional to  $\mathrm{Re}[n_\mathrm{RCP} - n_\mathrm{LCP}]$.
The same concept applies to the Kerr effect in reflection.
The Kerr- and Faraday rotations can therefore equally be regarded as \textit{circular birefringence} effects, i.e.\ a birefringence of circularly polarised light. The difference in absorption conversely causes different amplitudes of the two circular modes [Fig.\,\ref{Fig-Sketch}\,(b), (c)], which in transmission leads to \textit{circular dichroism} in \ref{Fig-Sketch}\,(c) and elliptically polarised light by superposition {in \ref{Fig-Sketch}\,(b), where the Faraday ellipticity 
is proportional to $\mathrm{Im} [n_\mathrm{RCP} - n_\mathrm{LCP}]$. Note that these refractive indices are different when $(\vec{k} \cdot \vec{m})
\neq 0$,  $\vec{k}$ being the wavevector of the light, {\red thus when there is a component of $m$ along the propagation direction.}

In Sect.\,\ref{Introduction} we have seen that in the case of the rotational effects a domain contrast is obtained by using an analyser and possibly a compensator in the reflection (or transmission) path that blocks the emerging amplitude differently for differently magnetised domains, thus causing a magnetisation-dependent intensity modulation of the reflected (or transmitted) light. 
In the case of pure {Kerr (or Faraday-) ellipticity [Fig.\ref{Fig-Sketch}\,(b)]}, a domain contrast cannot be generated by an analyser alone due to the elliptical character of the light (analyser \textit{and} compensator would be required to transform the two ellipses into a detectable rotation). Instead, one can exploit the (normalized) MCD intensity difference [Fig.\ \ref{Fig-Sketch}\,(c)],
\begin{eqnarray}
\label{eqn:Kerrcontrast}
A_\mathrm{C} = \frac{I_\mathrm{RCP} - I_\mathrm{LCP}}{I_\mathrm{RCP} + I_\mathrm{LCP}}  \propto \mathrm{Im}[n_\mathrm{RCP} - n_\mathrm{LCP}] \,\,,
\end{eqnarray}
of the transmitted intensities $I_\mathrm{RCP}$ and $ I_\mathrm{LCP}$, which doesn't require an analyser or compensator.

For the transverse Kerr effect, $(\vec{k}\cdot\vec{m}) =0$ and hence the refractive indices do not depend on $m$ in linear order. The effect instead results in the {\red afore}mentioned amplitude modulation, for which an analyser would as well be useless. 
This effect is therefore mainly applied in MOKE magnetometry for measuring purposes by using a photodetector \cite{Krinchik1968, McCord1995}. 
Also domain imaging under pure transverse conditions has been demonstrated by sequentially building up an image from laser-illuminated spots in a scanning Kerr microscope \cite{Buescher1993}. 
For wide-field Kerr microscopy, however, the transverse Kerr effect has so far only been applied \cite{Rave1987} as ``rotational" effect by polarising the light at 45$^\circ$ to the plane of incidence. Then the perpendicular light component is not affected by the magnetisation, while the parallel component receives an amplitude modulation leading to a detectable rotation in the superposition of the two components \cite{Dove1963}.

The reflection MO effects that appear as a magnetic intensity modulation can be described by the reflection matrix coefficients for $s$- and $p$-polarised light\begin{equation}
\vec{E}_\mathrm{out} = \mathbf{r} \cdot \vec{E}_\mathrm{in},
\end{equation}
with $\vec{E}_\mathrm{in}= (E_s, E_p)$.
The $2 \times 2$  matrix $\mathbf{r}$ depends on the magnetisation $\vec{m}$ and can be written  in Jones' notation
as 
\begin{eqnarray}
\mathbf{r} = \left( \begin{array}{ccc}
	r_{ss} &\quad \Delta_{sp} \vspace{0.05in} \\
	\Delta_{ps} &\quad r_{pp}+\Delta_{pp} 
\end{array} \right).
\end{eqnarray}
Here $r_{ss}$ and $r_{pp}$ are the standard, nonmagnetic {(Fresnel)} reflection coefficients for $s$- and $p$-polarised light, whereas the coefficients $\Delta_{sp}$, $\Delta_{ps}$ and $\Delta_{pp}$ depend linearly on $\vec{m}$. One can utilize the three orthonormal directions of the geometry to define the transverse $m_\mathrm{T}$, longitudinal $m_\mathrm{L}$, and polar $m_\mathrm{P}$ magnetisation components, respectively, 
such that $\vec{m} =(m_\mathrm{T}, m_\mathrm{L}, m_\mathrm{P})$ [\onlinecite{Oppeneer2001}]. The coefficient $\Delta_{pp}$ depends on $m_\mathrm{T}$ only, whereas $\Delta_{sp}$ and $\Delta_{ps}$ depend on both 
$m_\mathrm{L}$ and $m_\mathrm{P}$ [\onlinecite{Yang1993}].

The transverse Kerr effect appears as a magnetic change in the reflected intensity caused by $\Delta_{pp}$. 
The corresponding asymmetry is given as 
\begin{eqnarray}
\label{eqn-AT}
A_\mathrm{T} = \frac{R_p (+{m}_\mathrm{T}) - R_p (-{m}_\mathrm{T})}{  R_p (+{m}_\mathrm{T}) + R_p (-{m}_\mathrm{T})} ,
\end{eqnarray}
with $R_p = |r_{pp} +\Delta_{pp}|^2$; using the expressions for $r_{pp}$ and $\Delta_{pp}$ (see Ref.\,[\onlinecite{Yang1993}]) it can be shown that $A_\mathrm{T} \propto Q_\mathrm{V}$ [\onlinecite{Oppeneer2001}]. It deserves to be mentioned at this point that $A_\mathrm{T}$ can equally well be measured with linearly \textit{and} circularly polarised [$\vec{E}_\mathrm{in} = \frac{1}{\sqrt{2}}(1, \pm i) E_0$] light, and even unpolarised light, as long as there is a nonzero $E_p$ component. 

For a longitudinal magnetisation it is possible to observe a dichroic reflectivity contrast under specific conditions.  First, using LCP and RCP radiation a magnetic asymmetry exists, given by \cite{Mertins2002}
\begin{eqnarray}
\label{eqn-AL}
A_\mathrm{L}^\mathrm{C} =  \frac{R_\mathrm{RCP} - R_\mathrm{LCP}}{  R_\mathrm{RCP} + R_\mathrm{LCP}}  ,
\end{eqnarray}
which is proportional to $\Delta_{sp} \propto Q_\mathrm{V}$. A consequence of the similarity with Eq.\,(\ref{eqn-AT}) 
is that, when the magnetisation vector is rotated in-plane from the transverse to the longitudinal direction, the contrast recorded with circularly polarised light will additively contain $A_\mathrm{T}$ and $A_\mathrm{L}$ and change goniometrically between these two. 

Second, using linearly polarised light it is possible to measure a longitudinal dichroic reflection contrast as well \cite{Berger1997,Oppeneer2003}. This rather unknown MO effect is maximal when the incoming light is plane-polarised at an angle $\theta$ of 45$^{\circ}$ with respect to the incidence plane,
but it exists also for other angles $\theta$.
It becomes maximal when the magnetisation direction is reversed in an applied field, or when measuring antiparallel domains, 
and is expressed as
$A_\mathrm{L}^\mathrm{L} = [R_\theta (+m_\mathrm{L}) - R_\theta (-{m}_\mathrm{L})]/ [ R_\theta (+{m}_\mathrm{L}) + R_\theta (-{m}_\mathrm{L})]. $
Explicit expressions were given previously and show that $A_\mathrm{L}^\mathrm{L} \propto Q_\mathrm{V} \sin 2\theta$ [\onlinecite{Oppeneer2003}]. 
Consequently, this provides a second option to measure magnetic intensity contrast using linearly polarised light. 
For convenient use, this dichroic contrast can be reformulated by using that $R_\theta (+{m}_\mathrm{L}) + R_\theta (-{m}_\mathrm{L}) \approx 2 R_\theta (0)  \approx 2 R_{-\theta} (0)$,
with $R_\theta (0)$ the reflectivity of the nonmagnetised material. Further, using that inverting the magnetisation $\vec{m}$ is equivalent with changing $\theta$ to $-\theta$ (see Ref.\,[\onlinecite{Oppeneer2003}]), one can rewrite $A_\mathrm{L}^\mathrm{L} $ as 
\begin{eqnarray}
A_\mathrm{L}^\mathrm{L}  \approx  \frac{ R_\theta (m_\mathrm{L}) - R_{\theta} (0)} {2 R_{\theta} (0)} -
 \frac{ R_{-\theta} (m_\mathrm{L}) - R_{-\theta} (0)}{2 R_{-\theta} (0)}.
 \label{eqn-ALL}
\end{eqnarray}
In this MO effect a maximum contrast appears for longitudinal domains at oblique incidence when contrasts for $\theta = 45^{\circ}$ and $-45^{\circ}$ are subtracted. Even without subtraction, each of the terms in Eq.\,(\ref{eqn-ALL}) 
already provides a contrast, but with opposite sign. At normal incidence the longitudinal contrast vanishes.
It deserves to be mentioned that for a \textit{polar} out-of-plane magnetisation an equivalent magnetic dichroic effect $A_{\rm L}^{\rm P}$ in reflection exists \cite{Oppeneer2003}
that {should thus as well be suitable} for magnetic imaging of such domains. Alternatively, the MCD in reflection can be used to image out-of-plane magnetisations.

To summarize, using plane-polarised light one can thus detect magnetic contrast as an amplitude modulation, using either
$A_\mathrm{T}$,  $A_\mathrm{L}^\mathrm{L}$, or $A_{\mathrm{L}}^{\mathrm{P}}$, depending on the direction of the domain magnetisation. The magnitude of contrast
will vary with the angle between the light's polarisation and incidence
plane and also with the direction of the magnetisation with respect to the principal orthonormal directions (i.e., polar, longitudinal, and transverse).
A further option is to use  circularly polarised light and the MCD in reflection that can be employed to image in-plane and out-of-plane magnetic domains when there is a component of the magnetisation along the wavevector, i.e.\ $(\vec{k} \cdot \vec{m}) \neq 0$.

At the end of this paper we will show that besides the so far discussed MO effects that are all linear in the magnetisation, also the (quadratic) Voigt effect can be employed for analyser-free domain imaging. This effect, which can be applied in transmission \cite{Dillon1958} as well as reflection geometry \cite{Schaefer1990} and which was also found in the X-ray regime \cite{Valencia2010}, occurs when the light propagates transverse to the magnetisation vector. 
Here the eigenmodes are two \textit{linearly} polarised waves with vibrational planes along and perpendicular to $\vec{m}$. 
Incident light, plane-polarised along one of these two directions, will thus not alter its polarisation in the magnetic medium. 
If the polarisation plane is at an angle to $\vec{m}$, however, the polarisation state will be changed with the strongest effect occurring at an angle of 45$^\circ$. 
This is due to the fact that both linear eigenmodes are experiencing different refractive indices $n_{\|}$ and $n_{\mathrm{\perp}}$ so that they proceed in the medium with different velocities and with different attenuations. The light thus experiences a magnetic linear birefringence proportional to $\mathrm{Re} [{n_{\|}-n_{\mathrm{\perp}}}] $ and a magnetic linear dichroism proportional to $\mathrm{Im}[{n_{\|}-n_{\mathrm{\perp}}}]$ (note that here the word `linear' refers to the polarisation mode of the light and not to the order of the effect). 
In case of linear birefringence, the two partial waves are retarded relative to each other so that the outgoing light is elliptically polarised with a handedness that depends on the relative orientation of the polarisation plane and magnetisation axis. Oppositely magnetised domains, following the same axis, can therefore not be distinguished in their birefringence effect. Magnetic Linear dichroism (MLD) results in a rotation of the emerging light due to the different amplitudes of the partial wave. Note that the phenomenology is thus {\red \textit{opposite}} to that of magnetic circular birefringence and dichroism, where birefringence causes a rotation and dichroism leads to ellipticity
(compare Fig.\,\ref{Fig-Sketch}) \cite{Mertins2005}.

While conventionally the ellipticity of light due to the Voigt effect is transformed into linearly polarised light by a compensator and then brought to a contrast with the help on an analyser \cite{Kuch2015}, it can be expected that the selective amplitude modulation due to the MLD effect can directly lead to a MO contrast as will be demonstrated in Sect.\,\ref{Voigt Effect}.


\section{Experimental}
\label{Experimental}

Most of the presented domain images have been obtained in a wide-field optical polarisation microscope (Carl Zeiss AxioScope) with K\"{o}hler illumination scheme \cite{McCord2015} [Fig.\,\ref{Fig-Microscope}(a)]. The light of four light-emitting diodes (LEDs) is guided to the lamp house by fibre optics \cite{Hofe2013, Soldatov2017a}. The ends of the fibres are physically located at a plane that is confocal with the back-focal plane of the objective lens.
 {By a beam splitter, consisting of a semipermeable mirror at an orientation of 45$^\circ$ to the incoming ray path, the light is deflected downwards to the objective lens [Fig.\,\ref{Fig-Microscope}(c)]. After reflection from the sample, the light passes the mirror again on its way towards the camera or oculars.}
A centered fibre would result in perpendicular incidence of light, while off-centered fibres lead to oblique incidence with two orthogonal planes that can be chosen by activating one of the four LEDs
{(in Fig.\,\ref{Fig-Microscope}(c) this is illustrated for LED-1 --- the angle of incidence was around 30$^\circ$ for all four LEDs)}. 
By alternately pulsing two LEDs synchronously with the camera exposure, two Kerr modes can quasi-simultaneously be run and the corresponding images be displayed \cite{Soldatov2017a, Marko2015}. 
Note that in the beam splitter only plane-polarised light along the $x$ and $y$ axes is reflected linearly towards the sample --- for other orientations the light hitting the sample is elliptically polarised.
{Either white, red (640\,nm) or blue (450\,nm) LED light was used as indicated in the figure captions.
A 20x/0.5 (magnification / numerical aperture) objective lens was used for all experiments in the wide-field microscope.}

Circularly polarised light was generated by using a rotatable linear foil polariser followed by a quarter-wave plate [Fig.\,\ref{Fig-Microscope}(b)]
that was placed right on the entry side of the beam splitter.
The fast axis of the quarter-wave plate was aligned diagonally at 45$^\circ$ to the $x$- and $y$-axes. Setting the polarisation plane {then} along the $x$- and $y$-axis results in RCP and LCP polarised light, respectively. 
{Two quarter wave plates were applied: One plate, designed for a wavelength of 635\,nm, was used for investigations with red LED light, while another one, designed for a wavelength of 550\,nm (green light), was employed for white and blue light \footnote{{The 635\,nm quarter wave plate is made from mica and was obtained from Bernhard Halle Nachfl.\ GmbH, Berlin, and the 550\,nm plate was actually a quarter wave compensator from Carl Zeiss AG, Jena.}}. For the second case it is  expected that the light is strongly elliptically polarised rather than circular.
}
Unless mentioned otherwise, imaging was performed in reflection geometry without an analyser.  
To prevent overexposure of the camera, the intensity of the LEDs (which under regular Kerr conditions need to be maximised) had to be dimmed by a grey-filter in the illumination path.

\begin{figure}
\center
\includegraphics[width=1\linewidth,clip=true]{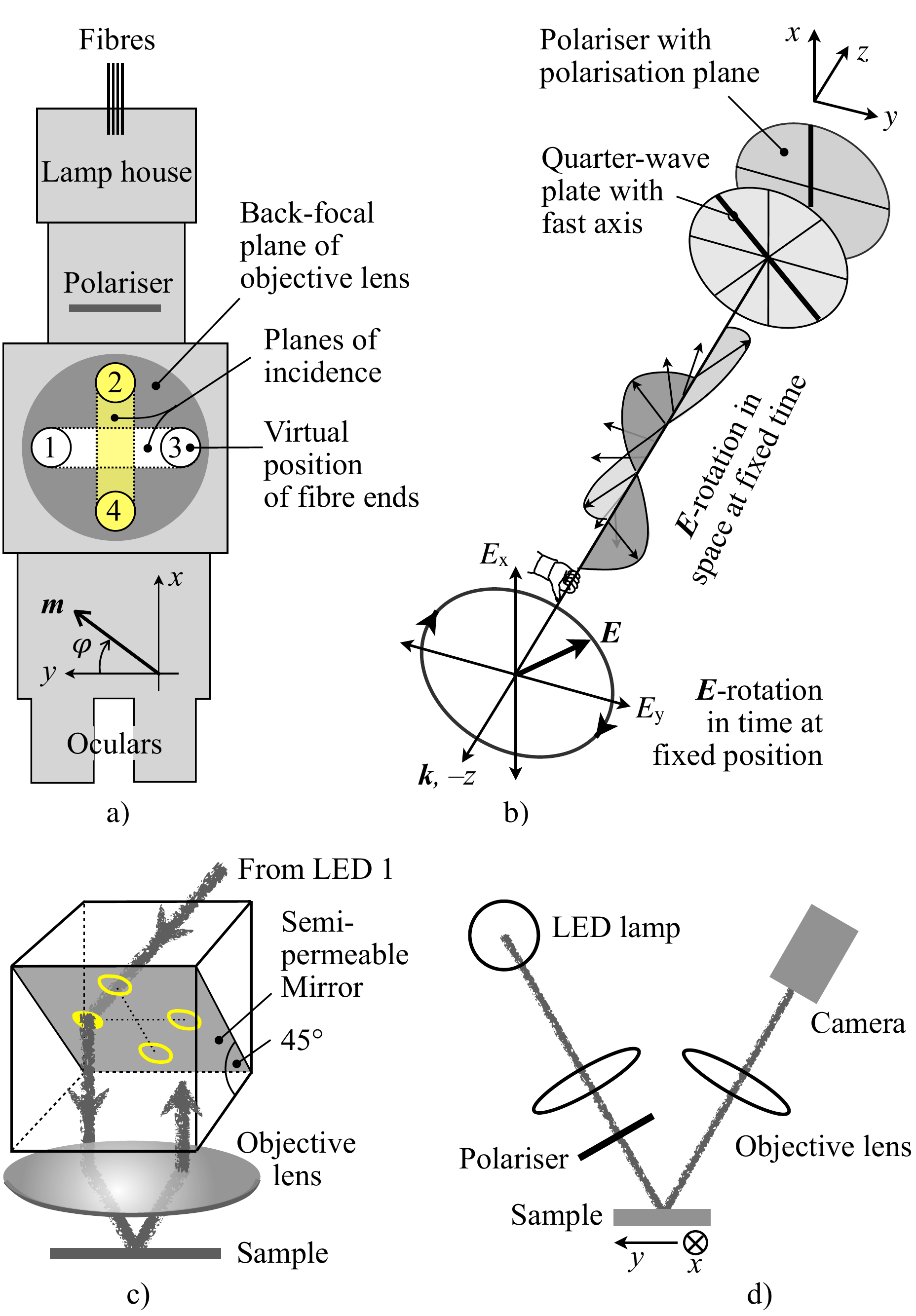}
\caption{(\textbf{a}) Schematic top view of our wide-field Kerr microscope, emphasising the two orthogonal planes of incidence that can be chosen by the activation of proper LEDs. The polariser, sitting in the illumination path, is by standard linear but can optionally be replaced by a circular polariser. The analyser (not shown) is located behind the objective on the reflection path. Indicated is the coordinate system to define the plane of incidence and the magnetisation direction. In the upcoming figures we will refer to this coordinate system and the given LED numbers. (\textbf{b}) Generation of (right-handed) circularly polarised light by a linear polariser and a quarter wave plate. Rotating the polariser by 90$^\circ$ would result in left-handed circular light. 
(\textbf{c}) Semipermeable reflector plate (beam splitter), sitting in the illumination path above the objective lens. Illustrated is the ray path for light from LED-1.  (\textbf{d}) Schematics of a simple Kerr microscope with separated illumination and reflection paths, which we have optionally used for selected experiments.}
\label{Fig-Microscope}
\end{figure}

Most experiments were preformed on an 
ideally oriented grain of a Fe-3wt$\%$-Si sheet with cube texture being characterized by a (100) surface with two easy axes of anisotropy parallel to the surface \cite{Hubert1998}. 
The specimen was coated with a zink-sulfide antireflection layer to enhance the Kerr effect \cite{Kranz1963}. A circular piece of such a 0.5\,mm thick sheet with a diameter of 10\,mm was placed between the pole pieces of a quadrupole electromagnet that allows for a computer-controlled rotation of the field by 360$^\circ$.  ``Kerr-sensitivity" curves were measured by plotting the image intensity of a {freely selectable sample area} in the saturated sample state as a function of magnetic field direction. The field direction thus corresponds to the magnetisation direction defined by the angle $\varphi$, see Fig.\,\ref{Fig-Microscope}(a).
All sensitivity curves were recorded on the unprocessed microscope image with the background image intensity adjusted to the same value by tuning the LED intensity.
All domain images shown on the FeSi material are difference images, in which a background image of the saturated state is subtracted to digitally enhance the contrast \cite{Schmidt1985}. 

For polar Faraday studies, a magneto-optical indicator film (MOIF) with perpendicular anisotropy was used as sample. The MOIF, provided by Matesy GmbH \cite{Matesy}, consists of a magnetic garnet film deposited on a transparent gadolinium-gallium-garnet (GGG) substrate and covered by a thin mirror layer. Polarised light enters the MOIF, passes the substrate and the magnetic film, and is then reflected from the mirror thus passing the magnetic film twice. In the magnetic garnet film a Faraday rotation is induced, which depends on the local magnetisation of the film and that is actually doubled due to the double penetration and the non-reciprocity of the Faraday rotation. The garnet film of the MOIF stack is thus effectively studied in transmission in a reflection polarisation (Kerr) microscope. ``Real" polar \textit{Kerr} microscopy was applied to an ultrathin cobalt-iron-boron single layer film (0.9\,nm thick) and to a Pt (3\,nm)/Co (1\,nm)/Pt (3\,nm) film, both with perpendicular anisotropy and not covered by antireflection coatings.
{All investigated samples are collected in Table\,\ref{Table}, showing the wavelength of the light and 
possibly the quarter wave plate that have been used together with the figures in which images of the specimens appear throughout the paper.}


\begin{table}[th!]
\tiny

\caption{{Summary of investigated samples. Listed is the wavelength of the applied LED light, the wavelength for which the quarter-wave plate is optimised, and the numbers of the figures in which the specimens are imaged.}}
\label{Table}
\begin{tabular}{| l |c|c|c|c|c|c|c|c|c|}%
\hline
  \multicolumn{1}{| l |}{ \, Sample}   & \multicolumn{5}{ c |}{FeSi}  & \multicolumn{2}{ c |}{Garnet} & \multicolumn{1}{c|}{CoFeB} & \multicolumn{1}{c|}{Pt/Co/Pt}\\[0.1cm]
 \hline
  $\rm {Wavelength} \atop {in ~nm}$ & 640 & 640 & 640 & 450 & white & 640& 640& white & white \\[0.1cm]
\hline
 \,$\lambda /4~{\rm plate} \atop {\rm in ~nm}$  & --- & 635 & 550 & 550 & 550 & --- & 635 & 550 & 550 \\[0.1cm]
\hline
  $ \rm \! Figure \atop \,number $ & $4,\, 5,\, 7,\, 13 \atop  14,\,16,\,17$ & 8,\,9 & 14,\,16 & 16 & 16 & $10,\,11 \atop  14,\,18$ & 10,\,15 & 12 & 14 \\[0.1cm]
\hline
\end{tabular}
\end{table}

To check for eventualities, selected experiments have been performed in a home-made microscope with separated illumination and reflection paths [Fig.\,\ref{Fig-Microscope}(d)]. The lateral resolution of this setup is only of the order of 30\,{$\mu$}m \cite{Schaefer2019}, it only allows for oblique incidence at a fixed angle of 30$^\circ$, but it has the advantage that the polarised light \textit{directly} hits the specimen without the need of a beam splitter.


\section{Conventional Longitudinal Kerr Microscopy}
\label{Conventional Longitudinal Kerr Microscopy}

To better understand the peculiarities of the analyser-free contrast phenomena in our FeSi material, let us briefly recap the typical domain contrast and sensitivity curves in conventional, analyser-based longitudinal Kerr microscopy. For the experiments on this material, \textit{red} light was used as it provides best domain contrast --- we have measured that blue and white light only lead to maximum longitudinal contrasts of 46\% and 51\%, respectively, of that of red light. Such trend is expected from the Kerr spectra of iron \cite{Oppeneer2001}.
As the best (obtained) contrasts of all sensitivity curves in this paper are below the maximum longitudinal Kerr contrast in red light, we have used that contrast $C_\mathrm{{max}}^\mathrm{{long}}$ as reference.
For each sensitivity curve presented below, we have calculated the contrast by $C = (I_2-I_1)/(I_2+I_1)$ with $I_{2}$ and $I_{1}$ being the maximum and minimum intensities, respectively, when turning the magnetisation in the range of 360$^\circ$.
Throughout this paper, all given Kerr intensities are normalised to $C_\mathrm{max}^\mathrm{long}$, which allows to compare the ``strengths" of all reported effects.

\begin{figure}
\center
\includegraphics[width=1\linewidth,clip=true]{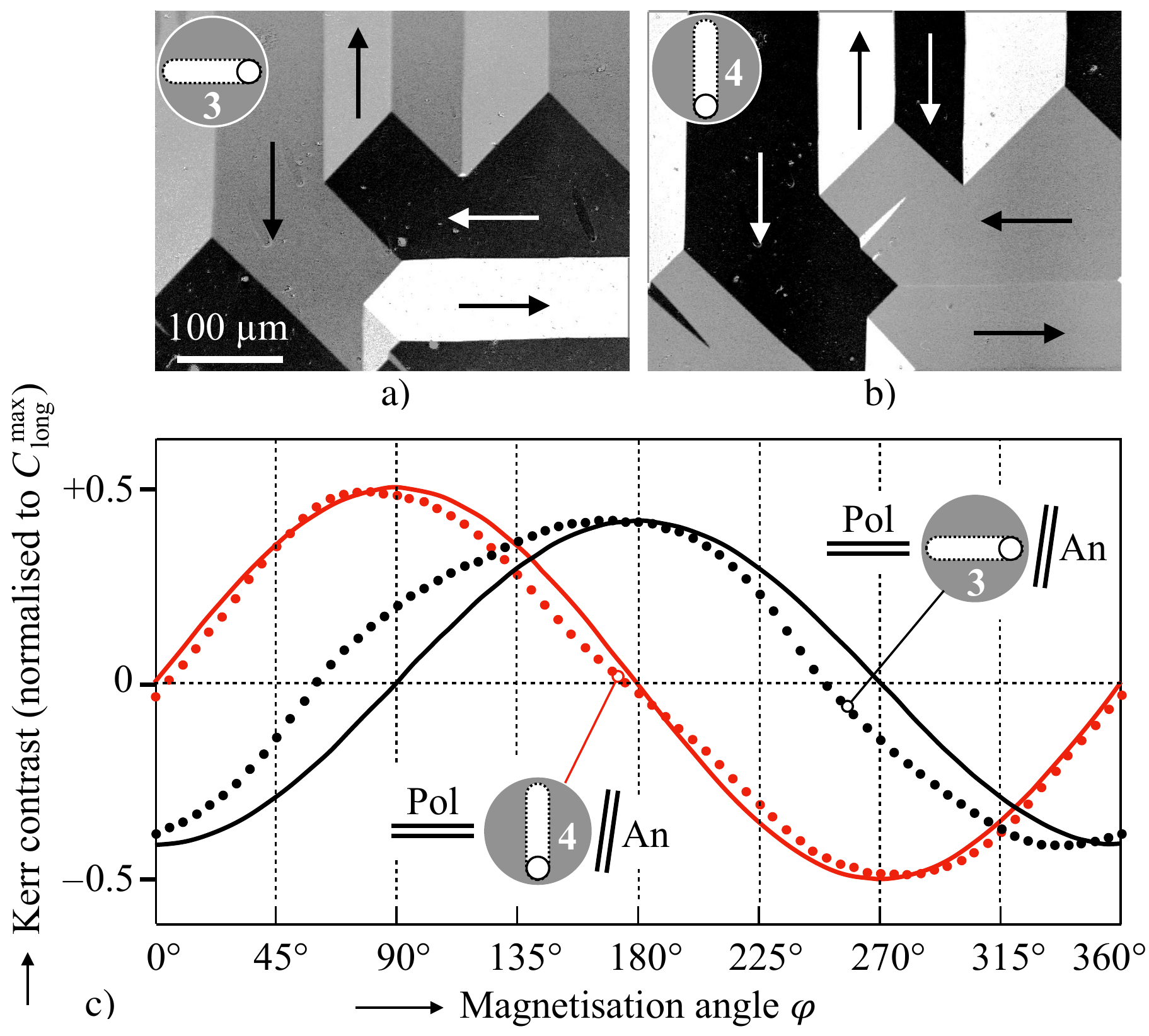}
\caption{Domains and sensitivity curves on our FeSi sample, recorded by the conventional longitudinal Kerr effect using an analyser in the reflection path of the microscope that was opened by an angle of 8$^\circ$ relative to the axis perpendicular to the polariser axis. Red LEDs at positions 3 and 4 were used and the polariser was along the $y$-axis, leading to $p$- and $s$-polarised light, respectively. The dotted curves are measured, while the full-line curves are regular sine and cosine functions to guide the eye.
Due to a superimposed transverse Kerr effect in case of $p$-polarisation, the two curves are phase-shifted by less than 90$^\circ$.
The arrows in the domain images indicate the magnetisation directions, which strictly follow the two orthogonal easy axes.
The maximum contrast of the sensitivity curves, shown in this plot, is used for normalising all intensities throughout the paper.
}
\label{Fig-Conventional}
\end{figure}

In Fig.\,\ref{Fig-Conventional} two complimentary domain images and sensitivity curves are shown, obtained with the red LEDs 3 and 4, i.e. at orthogonal planes of incidence. The polariser was fixed along the $y$-axis in both cases. So, the light from LED-3 hits the sample in the $p$-polarised state, whereas it is $s$-polarised for LED-4 corresponding to the two longitudinal cases in Fig.\,\ref{Fig-Kerreffects}.
While the domain contrast transverse to the plane of incidence disappears for LED-4 as expected for longitudinal Kerr sensitivity [Fig.\,\ref{Fig-Conventional}(b)], there is considerable transverse contrast for LED-3 [Fig.\,\ref{Fig-Conventional}(a)]. 
This is also visible in a notable phase shift of the corresponding sensitivity curve in Fig.\,\ref{Fig-Conventional}(c). 
Here also $\sin\varphi$ and $-\cos\varphi$ functions are plotted that should ideally be measured under \textit{pure} longitudinal sensitivity conditions to better see the apparent deviations between measured and ideal curves.
In Ref.\,[\onlinecite{Soldatov2021}] we have thoroughly investigated this effect, proving that it is caused by an intensity modulation of the reflected light due to the transverse Kerr effect, which is superimposed on the longitudinal rotation effect that primarily determines the contrast. As explained in Fig.\,\ref{Fig-Kerreffects}, a transverse Kerr effect is only expected for $p$-polarised light in accordance with the findings in Fig.\,\ref{Fig-Conventional}.

Besides the phase shift of the sensitivity curve measured with $p$-polarisation, both curves in Fig.\,\ref{Fig-Conventional} are more or less distorted.
This indicates that besides the transverse Kerr effect also other intensity-modulation effects might be superimposed  --- in the remainder of this paper several such effects will be identified. 
Such curve distortions need also to be considered when sensitivity curves are used to calibrate the domain intensity for quantitative Kerr microscopy \cite{Rave1987, Soldatov2017c}. Just recording the intensity at field angles that are multiples of 45$^\circ$ and assuming that the intensity is a strictly linear function of the longitudinal and transverse magnetisation components (as suggested in Ref.\,[\onlinecite{Rave1987}]) may be dangerous. It is furthermore worth to be noticed that the curve distortions can be strongly reduced \cite{Soldatov2021} by running the microscope in the pure longitudinal mode \cite{Soldatov2017a}, i.e.\ by pulsing opposite LEDs in synchronisation with the camera and subtracting the two pictures.


\section{Transverse Kerr Microscopy}
\label{Transverse Microscopy}

\begin{figure}
\center
\includegraphics[width=1\linewidth,clip=true]{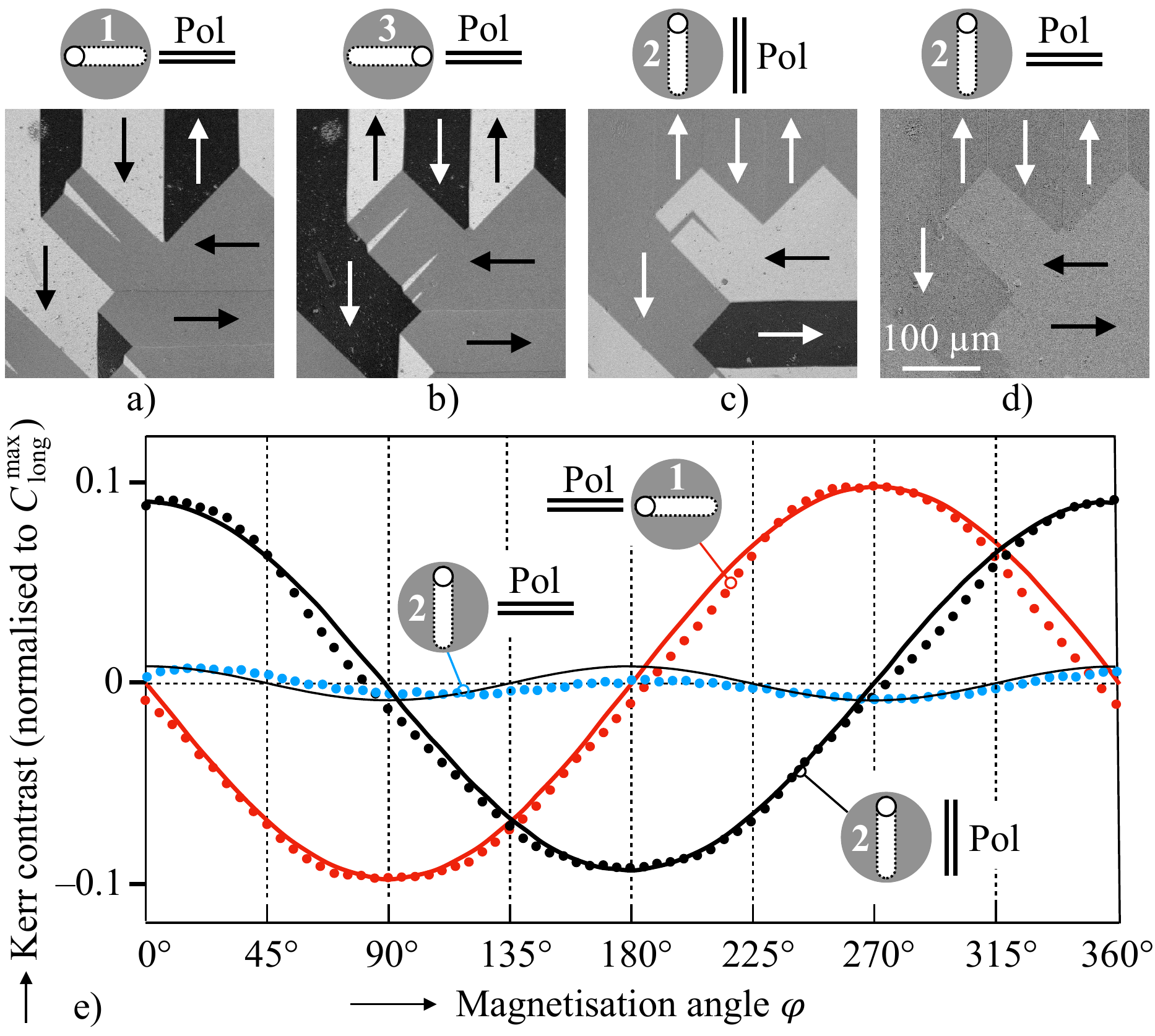}
\caption{
Analyser-free Kerr microscopy under pure transverse conditions, again by using red light. The domain images in (\textbf{a}) - (\textbf{d}) were obtained on the FeSi sheet under the indicated conditions [in (a), for instance, the sample was illuminated by LED-1 and the light was $p$-polarised, while for (b) the direction of incidence was inverted under the same polarisation]. The plots in (\textbf{e}) show the Kerr {\red sensitivities} for three cases. 
To guide the eye, $\cos\varphi$, $-\sin\varphi$, and $\cos2\varphi$ functions have been added as they are expected to be measured under ``ideal" conditions. 
The contrast is normalised to $C_\mathrm{max}^\mathrm{long}$, i.e.\ the maximum contrast that was achieved under conventional, longitudinal conditions (compare Fig.\,\ref{Fig-Conventional})}
\label{Fig-Transverse}
\end{figure}

In the previous section we have seen that the conventional longitudinal Kerr contrast, obtained by the use of an analyser, can be modulated by a superimposed transverse Kerr effect that leads to a phase shift of the otherwise longitudinal sensitivity curve.
By choosing $p$-polarised light and removing the analyser, the conditions are ready to measure the \textit{pure} transverse Kerr effect. An intensity modulation of the reflected light is then expected that depends on the magnetisation direction of an in-plane magnetised specimen (compare Fig.\,\ref{Fig-Kerreffects}). 
While \textit{quasi-transverse} imaging at 45$^\circ$ polarisation is well established as mentioned in Sect.\,\ref{Introduction}, wide-field Kerr microscopy under \textit{pure transverse} conditions has not yet been published to our best knowledge.
This is somewhat surprising as a pure transverse domain contrast can well be seen as demonstrated in Fig.\,\ref{Fig-Transverse}. Enhanced by background subtraction, the contrast is even of comparable strength as that of conventional, longitudinal Kerr microscopy.

The contrast symmetry corresponds to expectations: By comparing Fig.\,\ref{Fig-Transverse}(a) and (b), it is obvious that the domain contrast transverse to the plane of incidence is inverted when the direction of incidence is inverted due to an inverted cross product $\vec{m}\times\vec{E}_\mathrm{in}$ [see also Eq.\,(\ref{eqn-AT})]. 
This domain contrast disappears when the plane of incidence and the polariser are rotated by 90$^\circ$ [Fig.\,\ref{Fig-Transverse}(c)], consistent with the vanishing transverse Kerr effect $A_\mathrm{T}$ for a longitudinal magnetisation, and also the longitudinal effect $A_\mathrm{L}^\mathrm{L}$ vanishes for $\theta =0$.
Now the horizontal domains are transverse to the plane of incidence thus showing up with maximum contrast.
If the light is polarised orthogonal to the plane of incidence [i.e.\ $s$-polarised, Fig.\,\ref{Fig-Transverse}(d)], no domain contrast would be expected at all as now the Kerr amplitude either vanishes for transverse magnetisation components, i.e.\ along the polariser axis, or a non-vanishing Kerr amplitude will appear for rotated magnetisation, which, however, leads to a longitudinal \textit{Kerr rotation} that cannot be detected without analyser. Nevertheless, a weak domain contrast appears in Fig.\,\ref{Fig-Transverse}(d) that is \textit{quadratic} in the magnetisation. It is characterised by antiparallel domains of the same colour and a contrast between domains magnetised at 90$^\circ$. 
In Sect.\,\ref{Voigt Effect} we will address this quadratic contrast phenomenon again, showing that it is caused by the MLD effect.
If the domains in Fig.\,\ref{Fig-Transverse}(d) would be imaged under the same conditions but with an analyser added, the vertical domains would show up with maximum contrast due to the aforementioned longitudinal Kerr rotation.

The sensitivity curves in Fig.\,\ref{Fig-Transverse}(e) confirm this contrast phenomenology. For $s$-polarised light a curve with low contrast of just 2\% of $C_\mathrm{max}^\mathrm{long}$ is measured. It can well be approximated by a quadratic $\cos2\varphi$ function. 
For $p$-polarisation, sinusoidal intensity-dependencies are measured with (approximate) zero crossings whenever the magnetisation is parallel to the plane of incidence. By turning the plane of incidence together with the polariser by 90$^\circ$ results in a phase-shift of the sensitivity function by approximately 90$^\circ$, i.e.\ only the component of magnetisation perpendicular to the incidence plane results in a variation of the reflected light intensity as expected for the transverse Kerr effect. 
Compared to the sensitivity curves in Fig.\,\ref{Fig-Conventional}, the maximum contrast of the transverse curves is only $20\%$ of $C_\mathrm{max}^\mathrm{long}$, but they are significantly less disturbed.


\section{Longitudinal 45$^{\circ}$-dichroic Kerr microscopy}
\label{Oppeneer Effect}

In Sect.\,\ref{Theory} we have introduced a longitudinal dichroic reflection effect that is expected to occur by using linearly polarised light like for the transverse Kerr effect. The longitudinal contrast should be maximum at a polarisation angle of 45$^{\circ}$ with respect to the plane of incidence. Under this condition, the incoming light has both, $s$- and $p$-polarisation components. For a longitudinal magnetisation, the reflected electric field components, $E_{\mathrm{out},s,p}$, of the $s$- and $p$-polarised components are altered by the magnetic reflection coefficient $\Delta_{sp}$, leading to a changed intensity of the reflected light. Note that when the magnetisation is rotated, the $p$-polarised component should cause a transverse Kerr effect as well that should be superimposed to this longitudinal dichroic effect.

The contrast phenomenology of the longitudinal dichroic effect can be visualised by applying the Lorentz concept like in Fig.\,\ref{Fig-Kerreffects}. In Fig.\,\ref{Fig-Oppeneer-model} this is illustrated for two polariser settings at $\pm45^{\circ}$ relative to the plane of incidence and a given longitudinal magnetisation direction. The $s$- and $p$-components of the incoming electric field vector, $E_{\mathrm{in},s,p}$, are responsible for the indicated Kerr vectors, $\vec{K}_{s,p}$, on the reflection paths. By vector addition with the normally reflected light vector, $\vec{N}$, this results in an extension [Fig.\,\ref{Fig-Oppeneer-model}{(a)}] or shortening (b) of the outgoing field vector $\vec{E}_\mathrm{out}$, i.e.\ an increase or decrease of the total reflected amplitude when the polariser is turned from $+45^{\circ}$ to $-45^{\circ}$. Turning the magnetisation by $+180^{\circ}$ (not shown) will lead to inverted Kerr vectors and thus to inverted reflected light amplitudes. The domain contrast will consequently change sign when turning the polariser from $+45^{\circ}$ to $-45^{\circ}$.

\begin{figure}
\center
\includegraphics[width=1\linewidth,clip=true]{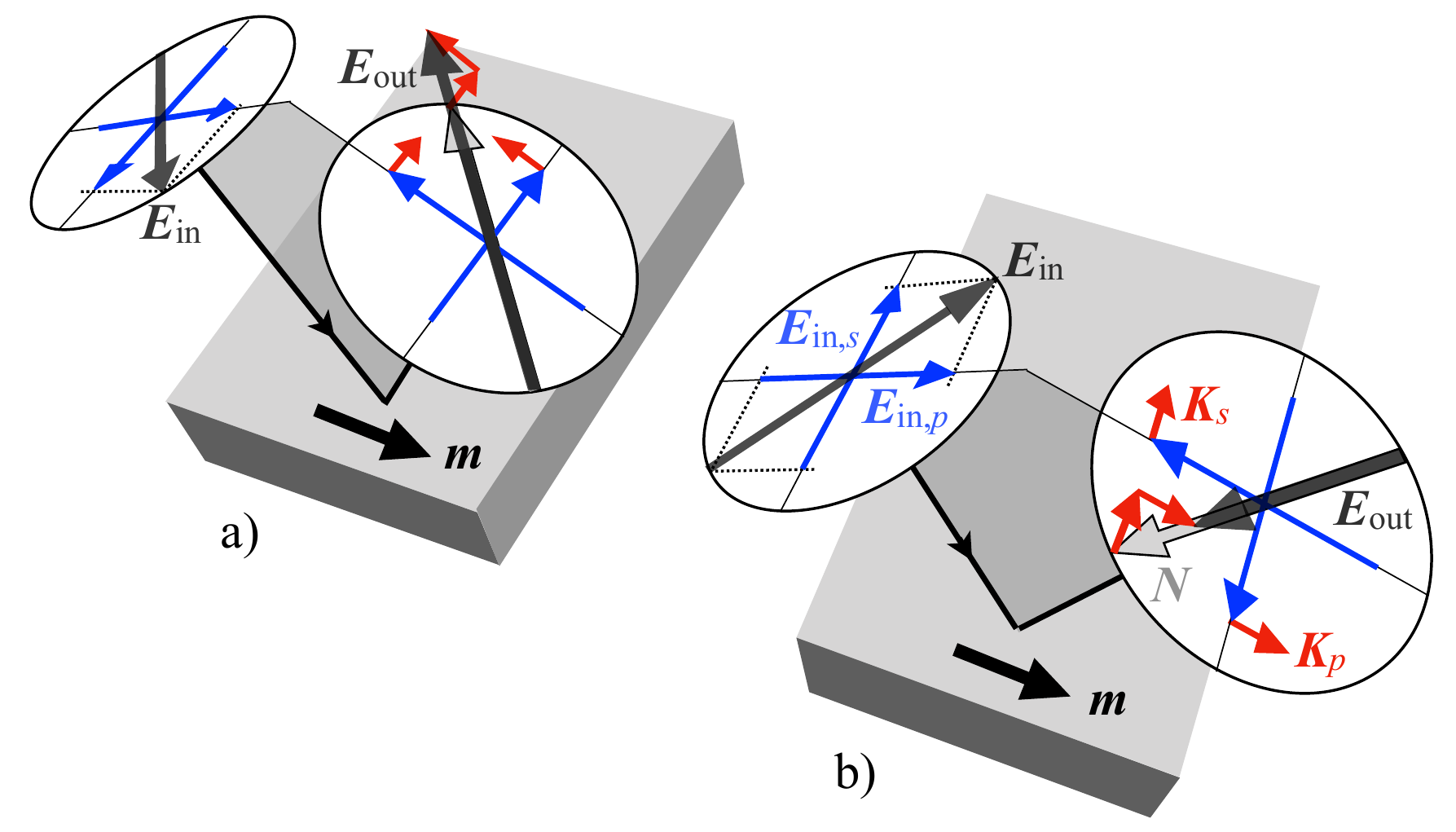}
\caption{
Lorentz concept of the longitudinal 45$^{\circ}$-dichroic Kerr effect. Shown are the electrical field vectors of the illuminating light ($\vec{E}_\mathrm{in}$), of the normally reflected light ($\vec{N}$), of the Kerr amplitude ($\vec{K}$), and of the totally reflected light ($\vec{E}_\mathrm{out}$). The $s$- and $p$-components of the incoming and normally reflected amplitudes are indicated by the blue vectors. The polariser, aligned at a $45^{\circ}$ angle relative to the plane of incidence in (a), is rotated by $90^{\circ}$ in (b). Compare with Fig.\,\ref{Fig-Kerreffects} for a similar illustration of the conventional Kerr effects.
}
\label{Fig-Oppeneer-model}
\end{figure}

In fact we were able to identify this predicted effect and its superposition with the transverse Kerr effect as demonstrated in Fig.\,\ref{Fig-Oppeneer}. Turning the polariser to $\pm45^{\circ}$ leads to sensitivity curves that are phase shifted by approx. $\pm45^{\circ}$ compared to those of the pure transverse or pure longitudinal effects (compare the curves in Fig.\,\ref{Fig-Oppeneer}(i) and (j) with those in Fig.\,\ref{Fig-Conventional} and Fig.\,\ref{Fig-Transverse}). This indicates that the transverse and the longitudinal 45$^{\circ}$-dichroic reflection effects are of about equal strengths. The maximum contrast is around 15\% of $C_\mathrm{max}^\mathrm{long}$. Compared to the pure transverse contrast it is thus reduced by a factor of approx. $\cos45^{\circ}$, which is conceivable when considering the 45$^{\circ}$ rotation of the light compared to pure transverse conditions. The domain images in Fig.\,\ref{Fig-Oppeneer}, obtained at polariser settings of $+45^{\circ}$ [images (a) to (d)] and $-45^{\circ}$ [images (e) to (h)] confirm the phenomenology of the sensitivity curves. Both, domains magnetised along as well as transverse to the plane of incidence show a comparable contrast as expected.

\begin{figure}
\center
\includegraphics[width=1\linewidth,clip=true]{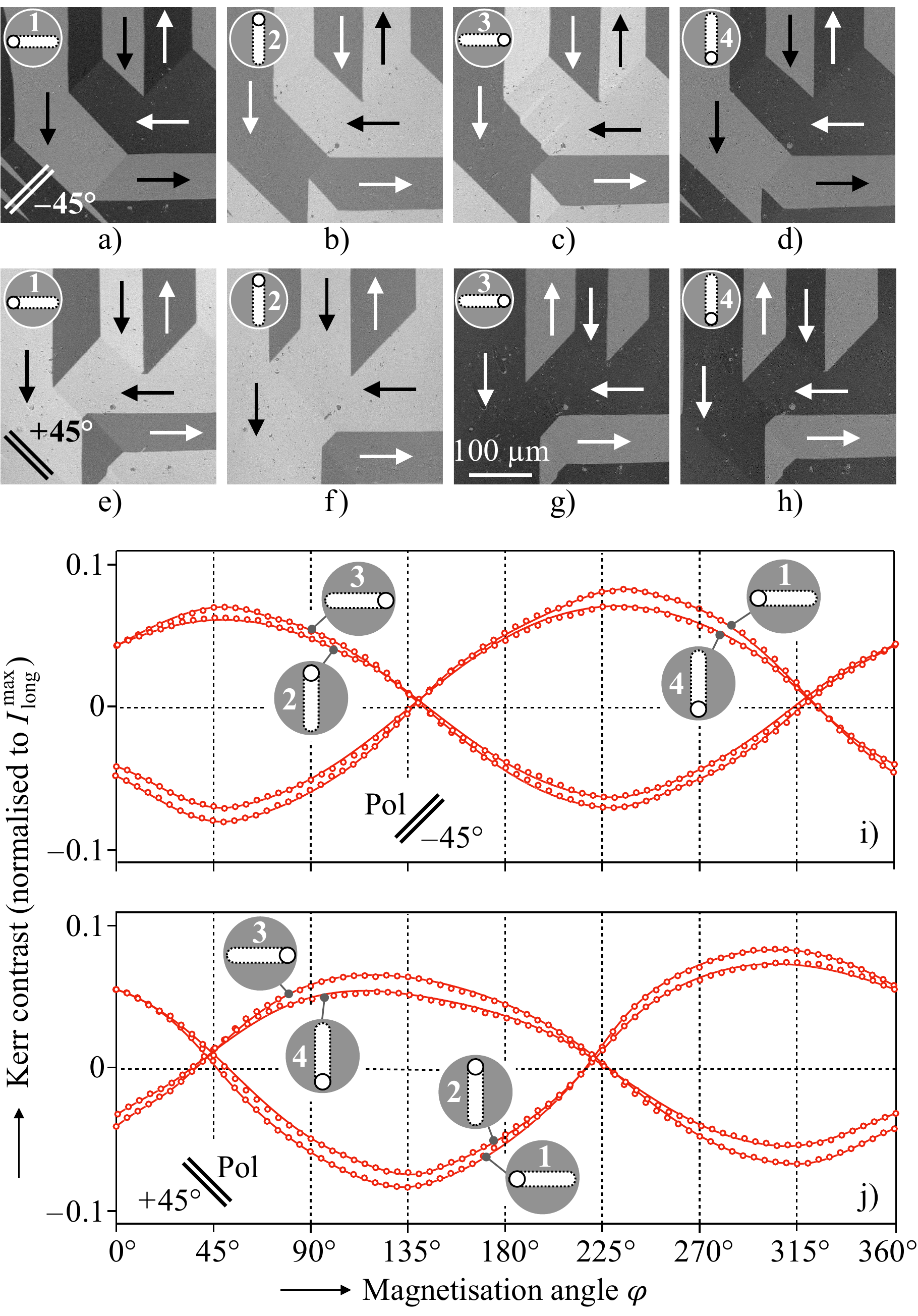}
\caption{
Analyser-free Kerr microscopy on the FeSi sheet by using red, \textit{linearly} polarised at angles of $\pm45^{\circ}$ relative to the planes of incidence. The domain images and the sensitivity curves were measured with all four LEDs thus comprising all possibilities for the plane and direction of incidence.
Domain images (a)--(d) were obtained at a polariser setting of $+45^{\circ}$, while the polariser was at $-45^{\circ}$ for images (e)--(h).
The contrast in (i),  (j) is again normalised to $C_\mathrm{max}^\mathrm{long}$.
}
\label{Fig-Oppeneer}
\end{figure}


\section{Longitudinal and transverse MCD-based Kerr microscopy}
\label{Longitudinal and transverse MCD-based Kerr microscopy}

In the recent article \cite{Kim2020}, {mentioned already in Sect.\,\ref{Introduction},} it was shown that the MO contrast of ultrathin magnetic films with perpendicular anisotropy can be significantly enhanced by embedding the magnetic media in between dielectric antireflection coatings, all deposited on a non-magnetic mirror film. That work appears
to be the first realization of a concept, which was suggested years ago \cite{Lissberger1961, Hubert1993, Wenzel1995}: by interference effects the MO amplitude in an ultrathin magnetic film can be enhanced while the regularly reflected light amplitude is reduced, leading to an increased MO rotation (the ZnS antireflection layer, applied to the FeSi sample in our paper, has a similar effect on bulk metallic specimens \cite{Kranz1963}).
Although the authors in Ref.\,[\onlinecite{Kim2020}] talk about MOKE microscopy, the domains were actually imaged by \textit{Faraday} microscopy as the magnetic films are optically transparent and as the light passes the films twice due to the mirror on the backside thus resembling Faraday microscopy in reflection geometry. 
The improved Faraday signal made it also possible to realise analyser-free MO microscopy by using \textit{circularly-polarised light} for illumination rather than plane-polarised light. 
By comparing Fig.\,\ref{Fig-Sketch}\,(a) and (c) it is evident that a Kerr- or Faraday rotation due to birefringence is not possible for circular illumination of a certain helicity, as for rotational effects both, left and right handed circular polarisation is required. 
In circular light, only the magnetic dichroism effect can rather be active 
as illustrated in Fig.\,\ref{Fig-Sketch}\,(c). It leads to differently damped amplitudes for different magnetisation directions that may result in a domain contrast.
In a wide-field Kerr (or Faraday-) microscope a domain contrast should thus be generated even in the absence of an analyser (like for the pure transverse Kerr effect, Sect.\,\ref{Transverse Microscopy}). In Ref.\,[\onlinecite{Kim2020}] this possibility of analyser-free imaging was {obviously} demonstrated for the first time.

Encouraged by that work, we have replaced the linear polariser by a circular polariser, removed the analyser and dimmed the light in our wide-field Kerr microscope. As sample we have again chosen the in-plane magnetised {FeSi} sheet with antireflection coating \footnote{We have started with a coated sample as antireflection coatings seemed to be an important ingredient for circular dichroism according to Ref.\,[\onlinecite{Kim2020}]. In the meantime we have found that with background subtraction a similarly good contrast can as well be seen on un-coated FeSi specimens (not shown).} that was already studied by transverse Kerr microscopy (Sect.\,\ref{Transverse Microscopy}). As being a non-transparent, bulk specimen, such a sheet can only be imaged in pure \textit{Kerr} (i.e.\ reflection) geometry. 
Using circularly polarised light for illumination, a magnetic contrast for in-plane magnetisation is expected to be achieved in both, the transverse Kerr configuration  [see Eq.\ (\ref{eqn-AT})] 
and in the longitudinal configuration [Eq.\ (\ref{eqn-AL})].

In fact it turned out that a domain contrast can readily be seen after background subtraction. 
Two series of images are presented in Fig.\,\ref{Fig-Circular-domains}. The upper row was obtained with {left-, the lower row with right-circular light.}
A couple of observations are noteworthy: 
\begin{itemize}
\item 
The contrast is inverted when the direction of incidence is inverted --- compare, e.g., images (a) and (c) or (b) and (d). 
\item 
Inverting the helicity of the light can lead to both, a contrast inversion [compare, e.g., images (d) and (h)] or no inversion [like in images (a) and (e)]. 
\item
In all images, a domain contrast is seen for both, $180^\circ$ domains magnetised vertically as well as horizontally. So, there is always contrast along and transverse to the plane of incidence. This indicates the simultaneous presence of longitudinal and transverse Kerr sensitivity.
\item 
Different levels of domain contrast are seen. Domains transverse to the plane of incidence always show a stronger contrast than those magnetised along the incidence plane.
\end{itemize}

From the fact that the contrast of the vertical domains does not change sign in Fig.\,\ref{Fig-Circular-domains}(a) and (e), and also \ref{Fig-Circular-domains}(c) and (g), we can conclude that in the transverse configuration the intensity contrast stems only from the $p$-polarised component of the circular light, i.e., the handedness does not play a role.
On the other hand, the contrast {inversion of the vertical domains} in panels \ref{Fig-Circular-domains}(b) and (f), and (d) and (h) is due to $A_\mathrm{L}^\mathrm{C}$. Reversing the helicity reverses the contrast signal, as expected according to Eq.\,(\ref{eqn-AL}).

\begin{figure}
\center
\includegraphics[width=1\linewidth,clip=true]{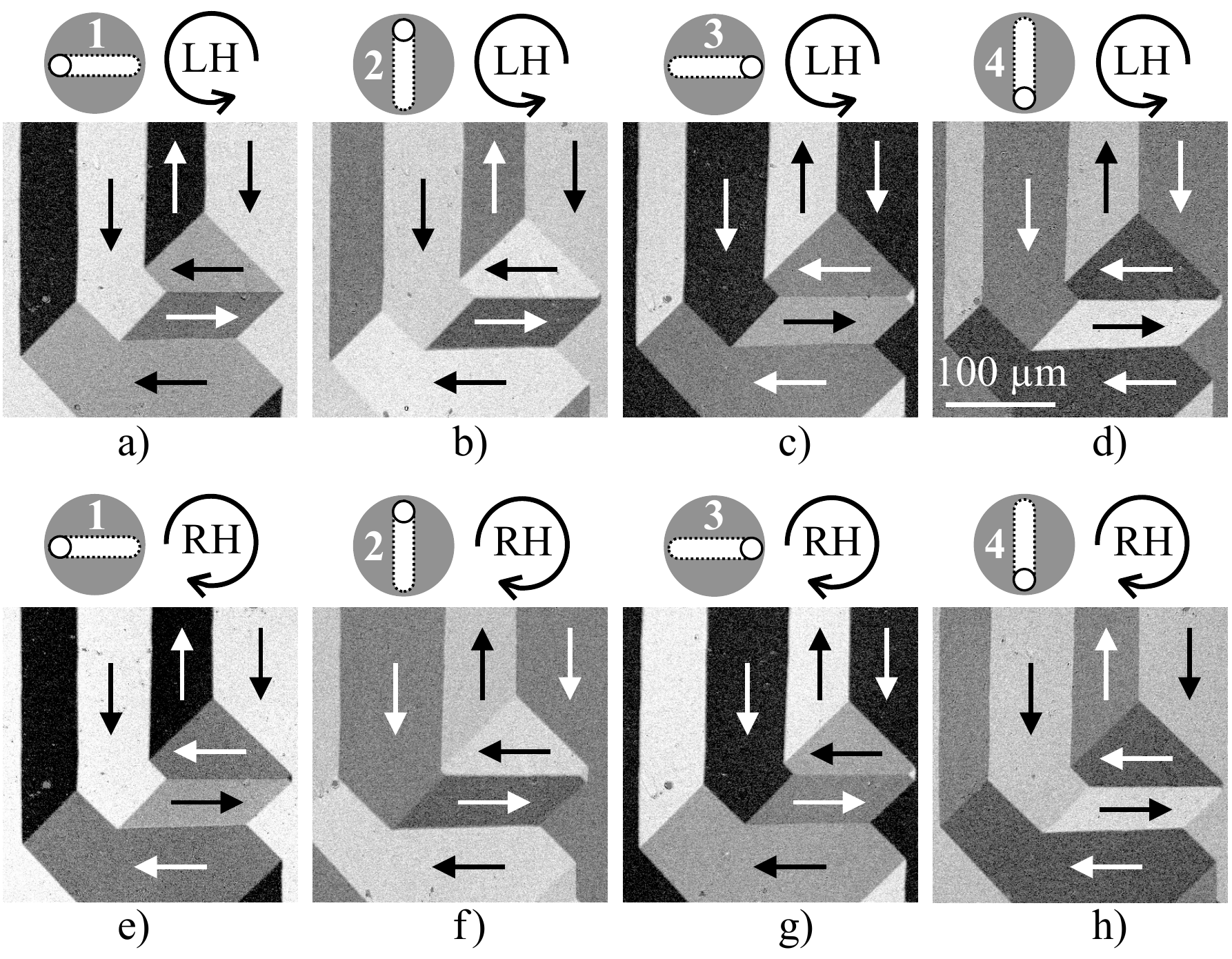}
\caption{Analyser-free domain imaging on the FeSi material by using red \textit{circular light} for illumination. 
{The 635\,nm quarter wave plate was applied to generate the circularly polarised light}. 
(\textbf{a}) - (\textbf{d}) Left-handed polarisation, (\textbf{e}) - (\textbf{h}) right-handed polarisation with all four LEDs separately activated as indicated. Shown are difference images of similar domain states.
}
\label{Fig-Circular-domains}
\end{figure}

The in-plane contrast symmetry under circular illumination is obviously different to that of the conventional, rotation-based longitudinal Kerr effect and the pure transverse Kerr effect.
This difference can readily be seen by comparing the sensitivity curves, obtained with circular light and displayed in Fig.\,\ref{Fig-Circular-curve}, with those of the conventional longitudinal (Fig.\,\ref{Fig-Conventional}) and {pure} transverse (Fig.\,\ref{Fig-Transverse}) Kerr effects. 
At orthogonal planes of incidence, the latter is characterized by $\sin\varphi$- and $\cos\varphi$-functions that are in-phase with intensity maxima and zero-crossings at magnetisation angles of (approx.) 0$^\circ$, 90$^\circ$, 180$^\circ$ and 270$^\circ$. 
In case of circular polarisation, the maxima and zero-crossings {are phase-shifted compared to pure transverse sensitivity curves. In the left column of Fig.\,\ref{Fig-Circular-curve} this is shown for left-handed and in the right column for right-handed circular light. 
To see the phase shift, also the cosine and sine functions have been added to the graph that would be expected under the (hypothetic) assumption of pure transverse sensitivities. Apparently, inverting the light helicity leads to an inversion of the phase shift direction. With light of the same helicity, the sensitivity curves are inverted for orthogonal planes of incidence [compare, e.g. curves (b) and (f) or curves (d) and (h)].}

The shapes of the sensitivity curves stem from the fact that with circular light a combination of both $A_\mathrm{T}$ and 
$A_\mathrm{L}^\mathrm{C}$ is measured simultaneously. 
For illustration, let us have a closer look at Fig.\,\ref{Fig-Circular-curve}(e) that shows the sensitivity curve for LED-2 and left-handed light.
{For $\varphi = 90^{\circ}$, $A_\mathrm{T} =0$, but $A_\mathrm{L}^\mathrm{C} \neq 0$. }
Near $45^{\circ}$ both, $A_\mathrm{T}$ and $A_\mathrm{L}^\mathrm{C}$ are nonzero; their contributions can cancel each other depending on the values of $\Delta_{pp}$ and $\Delta_{sp}$ [in curve (e) cancellation occurs around $\varphi = 55^{\circ}$]. 
At $\varphi = 0^{\circ}$, $A_\mathrm{T}$ will be largest while $A_\mathrm{L}^\mathrm{C} = 0$. 
At $\varphi = 135^{\circ}$ both $A_\mathrm{T}$ and $A_\mathrm{L}^\mathrm{C}$ are again nonzero, but now $A_\mathrm{L}^\mathrm{C}$ has a reversed sign, giving thus an additive enhancement of the contrast signal. 
Reversing the helicity [Fig.\,\ref{Fig-Circular-curve}(f)] will reverse the $A_\mathrm{L}^\mathrm{C}$ signal but not $A_\mathrm{T}$. At the angles $\varphi$ where the signals approximately cancelled in (e), they will now enhance each other, and \textit{vice versa}. 

According to Fig.\,\ref{Fig-Circular-curve}, the curve shifts may vary between some degrees and $45^{\circ}$. This can be due to the ``vectorial" mixing of signals in $|\vec{E}_{\rm{out}}|^2$, which is not simply an addition, i.e., $B_\mathrm{L} \cos \varphi \pm B_\mathrm{T} \sin \varphi$, of two modulated amplitudes  $B_\mathrm{L}$ and  $B_\mathrm{T}$. 
One can also see that the phase shifts are larger for LEDs\,2 and 4. These are the two LEDs for which the light is reflected orthogonally at the beam splitter mirror [see Fig.\,\ref{Fig-Microscope}(c)]. The beam splitter thus seems to have some influence on the light. We will address this again in Sect.\,\ref{Polar MCD-based Faraday microscopy}.

In any case,
the essential finding is that circularly polarised light causes a dominating transverse Kerr sensitivity (i.e. sensitivity to magnetisation components transverse to the plane of incidence) that is superimposed by a weaker longitudinal sensitivity (along the plane of incidence). In Fig.\,\ref{Fig-Circular-curve} we have drawn lines between the intensities at magnetisation angles at which deviations from zero-intensity are an indication for the existence of longitudinal contributions. The slopes of those curves immediately visualise the inversion of the longitudinal contrast on reversal of the helicity and direction of incidence. The overall maximum contrast, obtainable by using circularly polarised light, is around 15\% of $C_\mathrm{max}^\mathrm{long}$ and thus comparable to that of the longitudinal dichroic effect of Sect.\,\ref{Oppeneer Effect}.

\begin{figure}
\center
\includegraphics[width=1\linewidth,clip=true]{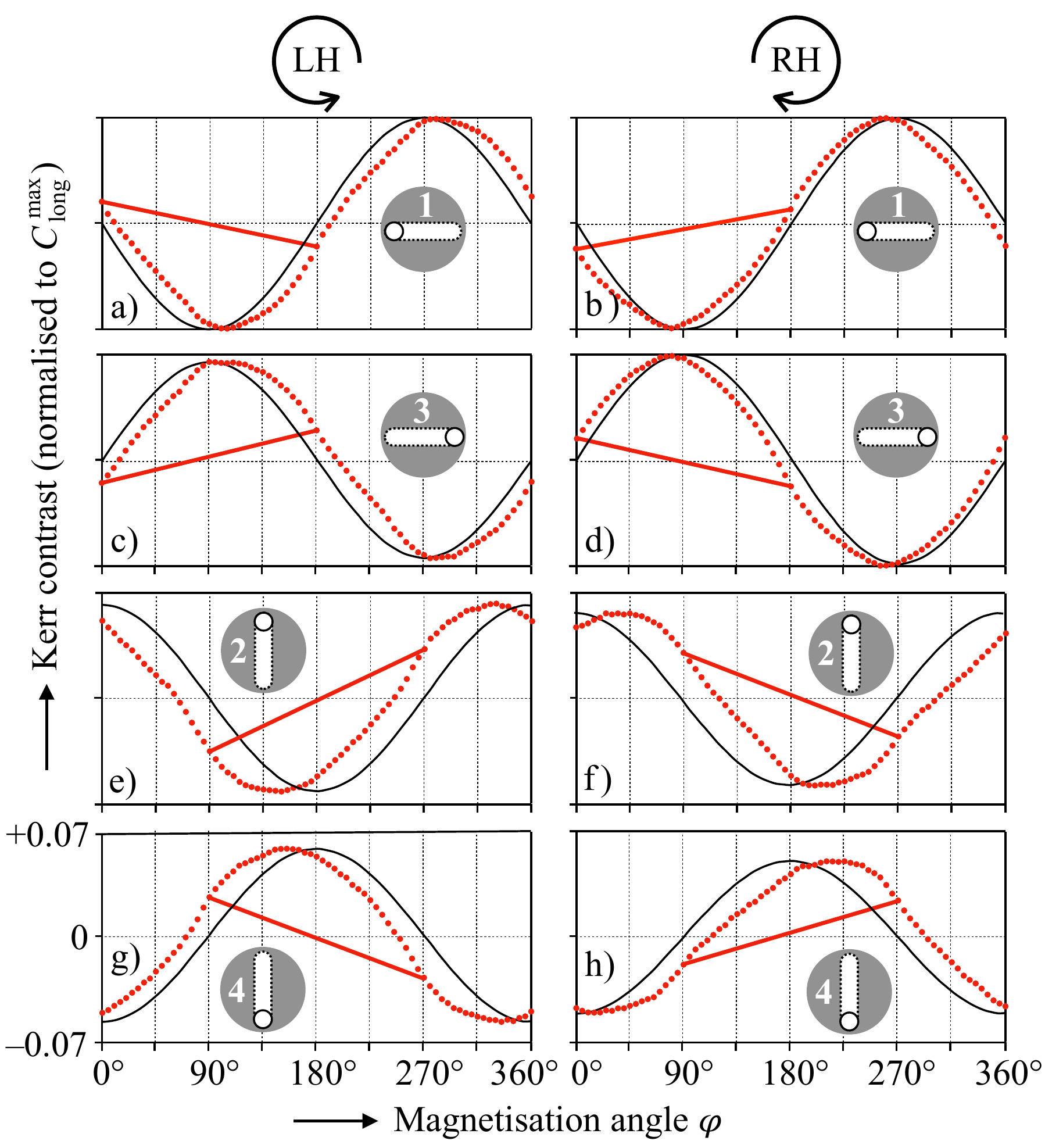}
\caption{
Sensitivity curves for orthogonal planes of incidence, measured on the FeSi sheet by using left-handed (left column) and right-handed (right column) circularly polarised, red light 
{like in Fig.\,\ref{Fig-Circular-domains}}. 
A non-vanishing slope of the straight lines indicates the presence of longitudinal contrast contributions. Also plotted are $\pm\cos\varphi$ and $\pm\sin\varphi$ functions that are supposed to indicate pure transverse sensitivity curves for a visual comparison.
}
\label{Fig-Circular-curve}
\end{figure}


{
\section{Polar MCD-based and 45$^{\circ}$-dichroic Faraday microscopy}
\label{Polar MCD-based Faraday microscopy}
}


The existence of a polar, MCD-based Faraday domain contrast, caused by the illumination with circular light, was already demonstrated in {Refs.\
[\onlinecite{Kim2020}] and [\onlinecite{Kuhlow1975}]} as mentioned in Sect.\,\ref{Introduction}. In Fig.\,\ref{Fig-MOIF-circular} a similar experiment is presented, but now on the garnet film with perpendicular magnetisation. In circular light [Fig.\,\ref{Fig-MOIF-circular} (b), (c)] the domain contrast is reduced compared to the (conventional) Faraday contrast that is optimised with the analyser (a). Inverting the handedness of the circular light leads to an inversion of the domain contrast, as expected.
{It needs to be noted that the MCD-based domain contrast on this specific specimen significantly depends on the selected focal depth of the microscope (a much stronger contrast than that visible in Fig.\,\ref{Fig-MOIF-circular} (b) and (c) can be obtained by defocusing the transparent garnet film by some micrometers, not shown). }

\begin{figure}
\center
\includegraphics[width=1\linewidth,clip=true]{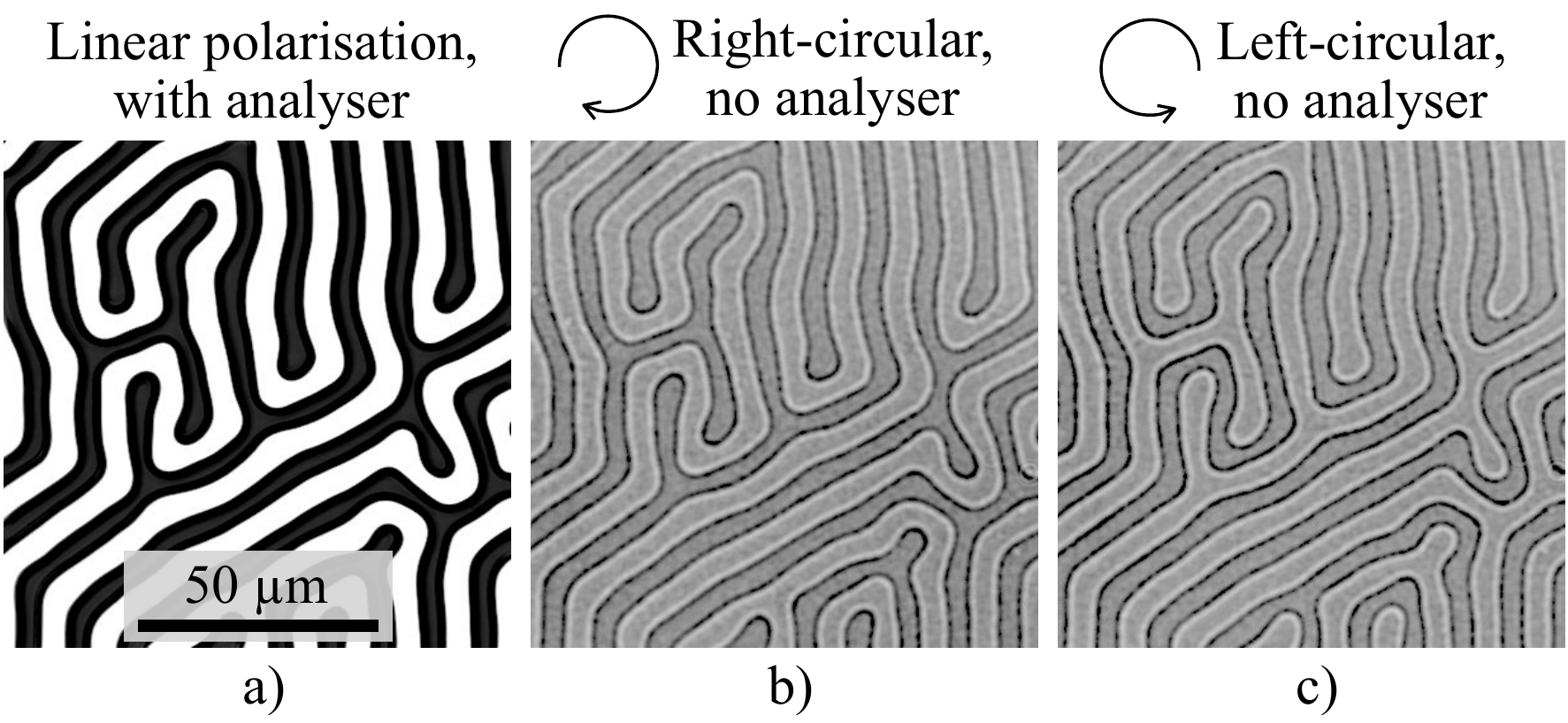}
\caption{Band domains {in our garnet film}, magnetised perpendicular to the film plane and imaged at perpendicular incidence. (\textbf{a}) Conventional polar Faraday microscopy using plane-polarised light and an analyser. In (\textbf{b}) and (\textbf{c}) right- and left-handed circularly polarised light was used without analyser. Red light was used, which in garnet material leads to slightly better contrast than white light (not shown). 
{Like in Figs.\,\ref{Fig-Circular-domains} and \ref{Fig-Circular-curve}, the 635\,nm quarter wave plate was applied to generate the circular light}. 
All images are live images without background subtraction.
}
\label{Fig-MOIF-circular}
\end{figure}

Interestingly, a domain contrast on the garnet film can also be seen by illumination with \textit{plane-polarised} light and omitting the analyser (Fig.\,\ref{Fig-MOIF-linear}). It has a curious symmetry, being maximal at polariser settings of $\pm45^\circ$ with inverted contrast. 
It thus resembles the phenomenology of the longitudinal dichroic Kerr contrast found on illuminating in-plane magnetised material with plane-polarised light (Sect.\,\ref{Oppeneer Effect}). {
It looks conceivable to address this contrast to the equivalent \emph{polar dichroic effect}, $A_{\rm L}^{\rm P}$, which was mentioned in Sect.\,\ref{Theory}.
There is, however, also a further effect that should lead to a domain contrast with the same phenomenology and which arises from the reflector module [Fig.\,\ref{Fig-Microscope}\,(c)] in our regular wide-field microscope:
At polariser settings along the $x$- and $y$-axes the light is mirrored {towards the sample} by keeping its polarisation direction and linear character. Linearly polarised light, falling on a specimen with up and down magnetised domains along the propagation direction, will lead to circular birefringence and circular dichroism [see Fig.\,\ref{Fig-Sketch} (a) and (b)]. Without analyser, however, the rotation due to birefringence cannot be detected and the dichroism effect leads to non-detectable ellipticity anyway.
Therefore no domain contrast is seen in the images (a) and (c) of Fig.\,\ref{Fig-MOIF-linear}. If the polarisation plane deviates from the $x$- and $y$-axes, however, the $x$- and $y$ components of the light are mirrored with a phase shift, leading to elliptically polarised light. Ellipticity will be maximal at the polariser angle of $45^\circ$ and it will be inverted at $-45^\circ$. The elliptical waves can well induce a circular (better ``elliptical") dichroism effect, thus resulting in a domain contrast that is strongest at polariser settings of $\pm45^\circ$ and which disappears at $0^\circ$ and $90^\circ$ in agreement with Fig.\,\ref{Fig-MOIF-linear}. Experimentally it is difficult to uniquely assign the $45^\circ$-contrast to one of the two effects.
}

\begin{figure}
\center
\includegraphics[width=1\linewidth,clip=true]{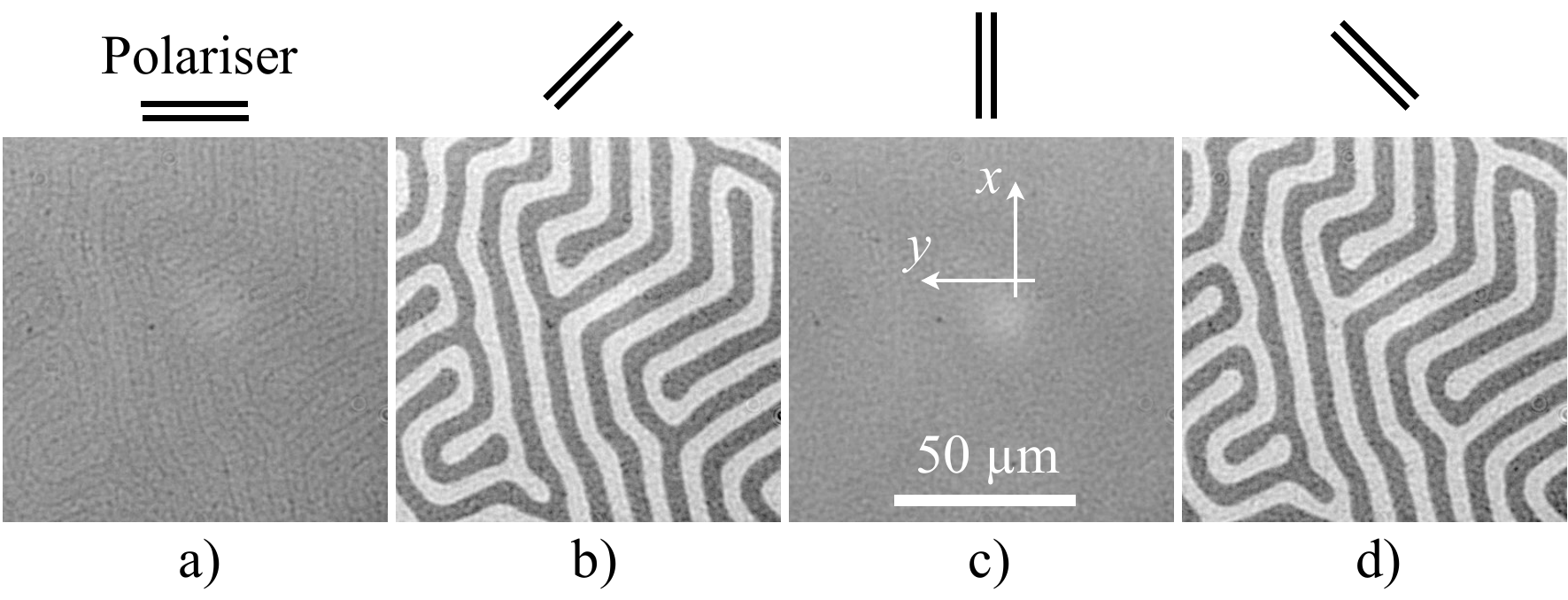}
\caption{Same band domains as in Fig.\,\ref{Fig-MOIF-circular}, again imaged at perpendicular incidence but now in \textit{plane-polarised} light at four settings of the polariser as indicated. Like in Fig.\,\ref{Fig-MOIF-circular}, the analyser was removed from the microscope.
}
\label{Fig-MOIF-linear}
\end{figure}

{
Although a domain contrast is not seen at polariser settings of 0$^{\circ}$ and 90$^{\circ}$ [Fig.\,\ref{Fig-MOIF-linear} (a) and (c)], the domain boundaries nevertheless show up with a dark line contrast as can be seen by a closer inspection of the two images. In (a) the line contrast is directly visible, while in (c) it only shows up after background subtraction (not shown). This domain boundary contrast is caused by magneto-optical diffraction. Let us postpone the discussion of such effects to Sect.\,\ref{Diffraction effects}.
}


{
\section{Polar MCD-based and 45$^{\circ}$-dichroic Kerr microscopy}
\label{Polar MCD-based Kerr microscopy}
}

Finally we have tested the encouraging results of Polar MCD-Based Faraday Microscopy (Sect.\,\ref{Polar MCD-based Faraday microscopy}) also on an ultrathin CoFeB metallic film with perpendicular anisotropy. Different to the mentioned experiments in Ref.\,[\onlinecite{Kim2020}], no mirror film was deposited underneath the magnetic film in our case so that it is justified to talk about true ``MOKE" microscopy. Furthermore we did not surround the magnetic film by interference layers. While by conventional polar Kerr microscopy a domain contrast can readily be seen without image processing [Fig.\,\ref{Fig-CoFeB}(a)] that can be infinitely enhanced by background subtraction (b), a contrast is hardly visible {in}  analyser-free Kerr microscopy in an un-processed image (not shown). After background subtraction, however, domains are well seen by using circular polarisation (c) as well as plane-polarised light (f). For the latter the polariser has to be set at $\pm45^\circ$ as elaborated in Fig.\,\ref{Fig-MOIF-linear}. In both cases, the strength of the contrast after background subtraction is comparable with that of an unprocessed image in the conventional polar Kerr mode [compare images (c) and (f) with image (a)].

This contrast can be further enhanced by activating two LEDs at opposite locations, running them in the pulsed mode, and subtracting the two corresponding images with proper normalisation --- a concept that was introduced in Ref.\,[\onlinecite{Soldatov2017a}] for the enhancement of contrast in conventional longitudinal Kerr microscopy.
In Fig.\,\ref{Fig-MOIF-circular} we have seen that the polar dichroic Faraday contrast is inverted by using LCP and RCP light. The same is true for the polar dichroic Kerr contrast as demonstrated in Fig.\,\ref{Fig-CoFeB}(c) and (d). This is the polar equivalent of the longitudinal MCD in reflection [Eq.\ (\ref{eqn-AL})]. 
Here we have placed two linear polarisers with polarisation axes along the $x$- and $y$-axes in the aperture plane of the microscope (which is confocal to the back-focal plane of the objective lens), one in the illumination path of LED-2 and the other in the path of LED-4, and both followed by a quarter-wave plate that is oriented at $45^\circ$. Consequently the light of LED-2 and LED-4 is right- and left-circularly polarised, respectively. Note that now we have oblique incidence of light at an angle of approx. $25^\circ$, which, however, has no significant influence on the polar Kerr effect that scales with the cosine of the angle of incidence. 
By running the two LEDs in the pulsed mode in synchronisation with the camera and by subtracting the two obtained images with inverted contrast, the contrast in the difference image is actually doubled after proper normalisation [see Fig.\,\ref{Fig-CoFeB}(e)]. 

The same concept can also be applied to analyser-free imaging in plane-polarised light [Fig.\,\ref{Fig-CoFeB}(f)-(g)]. Here the light of the two LEDs was linearly polarised at $\pm45^\circ$, in accordance with the polar dichroic reflection effect mentioned in Sect.\ \ref{Theory}.
The magnetic contrast is proportional to $Q_\mathrm{V} \sin 2\theta$, with $\theta$ the angle of the plane-polarisation with respect to the reflection plane.
This gives a maximum contrast for $\theta = 45^\circ$ and an inverted contrast for $135^\circ$, as seen in Fig.\,\ref{Fig-CoFeB}(f)-(g)]. Taking the difference image thus doubles here as well the contrast [cf.\ Eq.\ (\ref{eqn-ALL})]. 
As mentioned when discussing the garnet film 
{in the previous section}, the mechanism of the contrast formation needs to be scrutinised further.
Notwithstanding, the pulsed mode makes analyser-free, intensity-based Kerr microscopy applicable also to ultrathin magnetic films without the necessity of dielectric interference layers.

\begin{figure}
\center
\includegraphics[width=1\linewidth,clip=true]{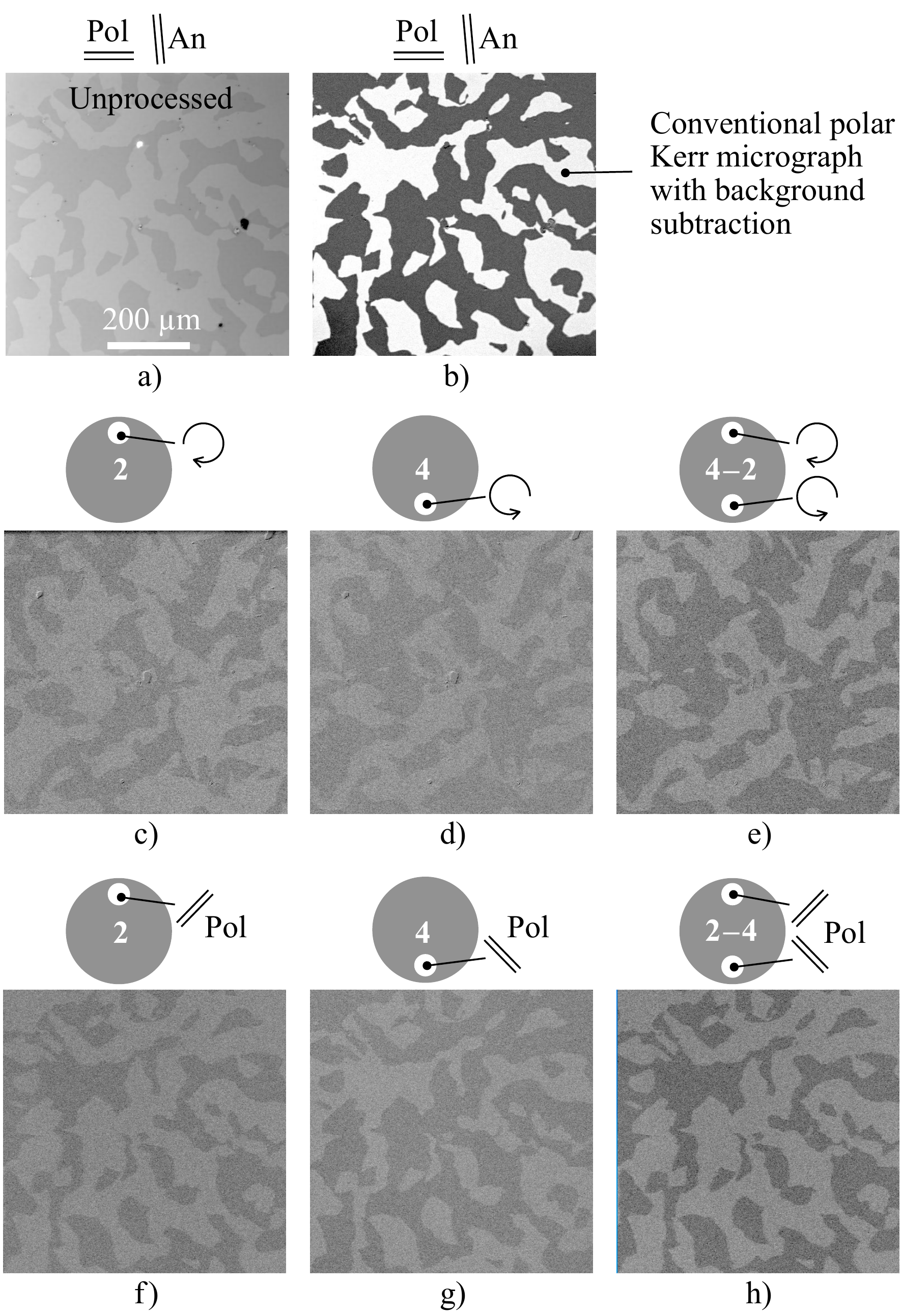}
\caption{Domains in Ta(3)/\textbf{CoFeB(0.9)}/MgO(1)/Ta(3) (thickness in nanometer) film with perpendicular anisotropy, imaged by Kerr microscopy under different conditions: (\textbf{a}) By conventional polar Kerr microscopy at perpendicular incidence. (\textbf{b}) Background subtraction significantly enhances the contrast. (\textbf{c}), (\textbf{d}), (\textbf{e}) Analyser-free imaging in circular {(better elliptical)} light and (\textbf{f}), (\textbf{g}), (\textbf{h}) by using plane-polarised light at oblique incidence (see text for details). White light was used for illumination {and for circular polarisation the 550\,nm quarter wave plate was applied.}}
\label{Fig-CoFeB}
\end{figure}

Those pulsed modes for contrast enhancement rely on the subtraction of two images with inverted contrast, obtained by choosing oppositely arranged LEDs for illumination. Consequently, they are only possible for oblique incidence of light. An interesting alternative for the case of circular polarisation, that should also work for perpendicular incidence, might be worth to be examined: In Ref.\,[\onlinecite{Narushima2016}] methods for (general) dichroism microscopy were suggested, which allow for a periodic and discrete alternation of the incident polarisation between left- and right-circularly polarised light by using a photoelastic modulator or a beam displayer combined with a chopper and quarter wave plate in the illumination path. Synchronising the alternating polarisation with the camera exposure and subtracting the corresponding two images should lead to the same kind of contrast enhancement as shown in Fig.\,\ref{Fig-MOIF-circular}.\\


\section{MLD-based Voigt- and Gradient Microscopy}
\label{Voigt Effect}

As introduced at the end of Sect.\,\ref{Theory}, the main difference of the Voigt contrast compared to the Kerr- (or Faraday) contrasts is its quadratic dependence on the magnetisation direction.
In conventional, analyser- and compensator based Voigt microscopy, an in-plane magnetised specimen is illuminated at perpendicular incidence thus suppressing the Kerr- (or Faraday) effect. In case of an Fe(100) surface, the polarisation plane needs to be at $45^\circ$ to the two easy axes to obtain maximum Voigt domain contrast \cite{Kuch2015}.

The experimental conditions are different when the MLD effect is applied directly in an analyser-free microscope. This is demonstrated in Fig.\,\ref{Fig-Voigteffect} for our FeSi sample. Perpendicular incidence of (red) light was chosen to avoid any of the other contrast phenomena discussed so far and the polariser was rotated in steps of $45^\circ$. Maximum domain contrast is found when the light is polarised along the two anisotropy axes, and this contrast is inverted by rotating the polariser by $90^\circ$ [compare Fig.\,\ref{Fig-Voigteffect}(a) and (b)]. This phenomenology clearly demonstrates that the different absorption of the two linearly polarised partial waves due to the MLD effect (see Sect.\,\ref{Theory}) is responsible for observed contrast: In our case, domains magnetised along the polarisation axis obviously lead to a stronger absorption and thus darker colour compared to those magnetised transverse to the polarisation axis. At polariser settings of $\pm45^\circ$ [Fig.\,\ref{Fig-Voigteffect}(c) and (d)] the domain contrast disappears as now the two orthogonal easy axes absorb the light equally. 

It should be noticed that the quadratic effect, measured in Fig.\,\ref{Fig-Transverse}(d) and (e) by illumination of the same sample with $s$-polarised light at \textit{oblique} incidence is as well caused by the MLD effect. Under those conditions the transverse Kerr effect is not possible and the longitudinal Kerr effect, being well possible though, leads to a light rotation that cannot be detected without analyser. The basic conditions with the polariser along one of the two easy axes are, however, suitable
for the MLD effect. Also the phase of the quadratic function in Fig.\,\ref{Fig-Transverse}(e) corresponds to expectations --- the contrast is maximal at magnetization angles along the easy axes rather than along diagonal direction, which would be true in analyser-based Voigt microscopy.

\begin{figure}
\center
\includegraphics[width=1\linewidth,clip=true]{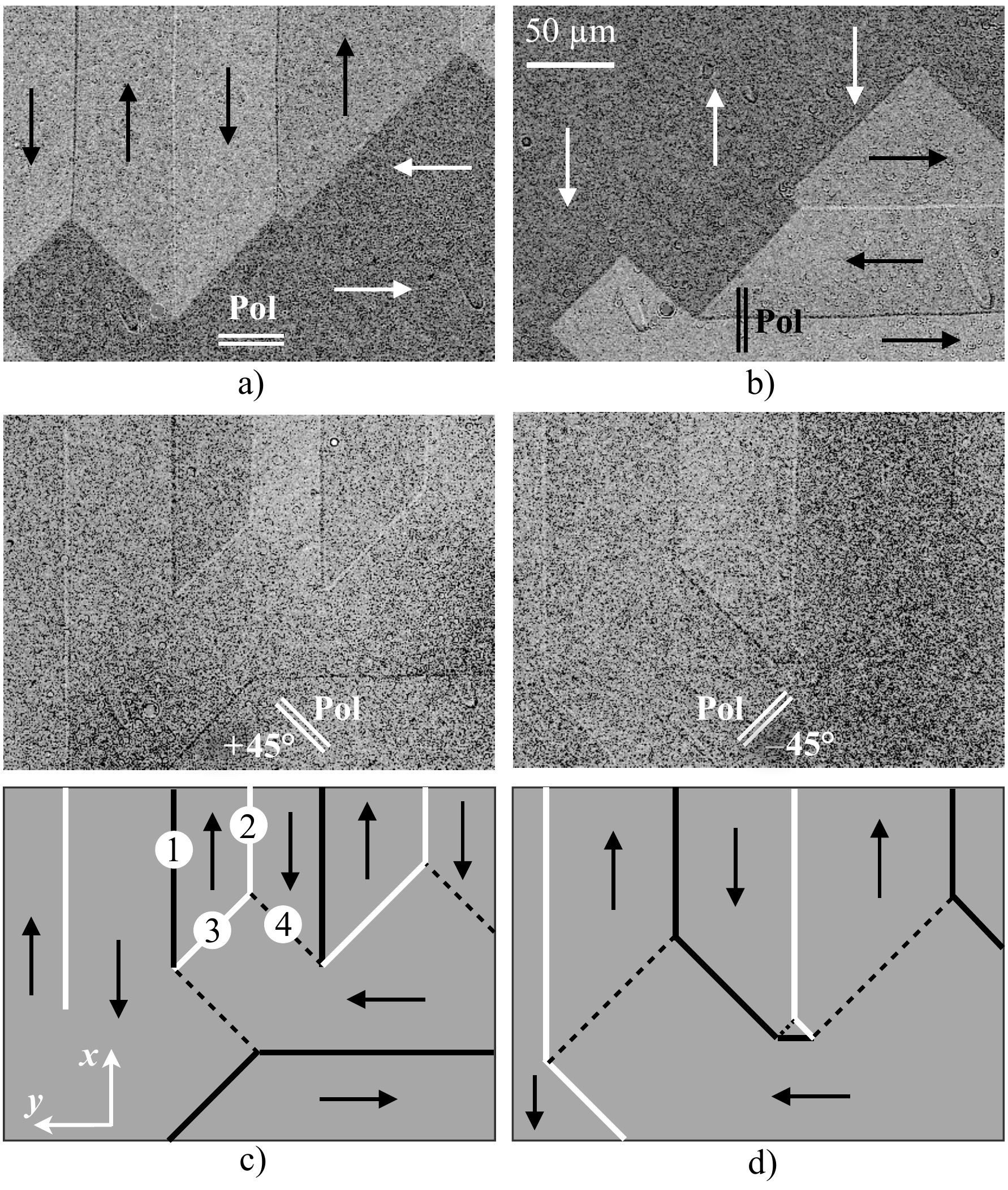}
\caption{
Domains on the FeSi sheet, imaged in red light at perpendicular incidence. The polariser was set along the $x$- and $y$-axis and at angles of $\pm45^\circ$ as indicated. {\red The contrast in (a) and (b) is due to the reflection MLD effect and the gradient effect, while (almost) pure gradient contrast is seen in (c) and (d).}
 The schematics in (c) and (d) trace the gradient contrast of the corresponding domain images.}
\label{Fig-Voigteffect}
\end{figure}

The domain boundary contrasts, visible in all images of Fig.\,\ref{Fig-Voigteffect}, are caused by the MO gradient effect \cite{Schaefer1990, Schaefer1990b, Kuch2015}. As mentioned in Sect.\,\ref{Introduction}, the gradient effect is sensitive to gradients in the magnetisation vector field which are strongest across domain walls. Already in the original work when this effect was discovered \cite{Schaefer1990} it was noticed that the effect occurs under the same experimental conditions as the Voigt effect and that it can be described phenomenologically by the dielectric law \cite{Schaefer1990, Thiaville1991}
\begin{eqnarray}
\label{eqn-GradientTensor}
\vec{D} = P_\mathrm{gr}\left( \begin{array}{ccc}
	-\frac{\partial m_x}{\partial y} -\frac{\partial m_y}{\partial x} &\quad \frac{\partial m_x}{\partial x} - \frac{\partial m_y}{\partial y} \vspace{0.05in} \\
	\frac{\partial m_x}{\partial x} - \frac{\partial m_y}{\partial y} &\quad \frac{\partial m_x}{\partial y} + \frac{\partial m_y}{\partial x}
\end{array} \right) \vec{E}_{\mathrm{in}}\, ,
\end{eqnarray}
in which the tensor contains various gradients of the in-plane magnetisation components. 
The material constant $P_\mathrm{gr}$ scales with that of the Kerr effect ($Q_\mathrm{V}$).
Like for (conventional) Voigt microscopy, also for gradient microscopy a compensator and analyser were needed for contrast adjustment, with the latter being aligned perpendicular to the polariser. The effect was therefore described by just considering the off-diagonal elements of the dielectric tensor --- an incoming wave, polarised along the $x$-direction for instance, will induce a MO component along the $y$-axis according to $D_{y} = \epsilon_{yx} E_{{\mathrm{in}},{x}}$ that is detected by the crossed analyser along the $y$ axis. At that time \cite{Schaefer1990} the diagonal elements of the tensor ($\epsilon_{xx}$ and $\epsilon_{yy}$) were only added for symmetry reason, pointing out that those components would just lead to an amplitude modulation of the reflected light that cannot be detected with the analyser. 

Without analyser, however, it is exactly those diagonal terms containing mixed derivatives that are responsible for the gradient contrast symmetry in the images of Fig.\,\ref{Fig-Voigteffect}. For a more detailed discussion we need to consider the magnetic microstructure of domain boundaries, an aspect that was discovered by Kambersk{\'y} \cite{Kambersky1992, Kambersky2011}. In Eq.\,(\ref{eqn-GradientTensor}) each of the tensor components contains sums of gradients rather than single gradients. Such sum-terms are necessary for specimens in which subsurface, perpendicular gradients are relevant to fulfil the condition div\,$\vec{m}=0$, which is e.g.\ the case for so-called V-lines in an iron-like material \cite{Hubert1998}. In our case we can neglect subsurface gradient components if we assume that the light --- within the information depth ---  does not interact with the internal Bloch-component of the stray-field free vortex domain walls \cite{Hubert1998}.
A strict in-plane magnetisation can then be assumed, the wall magnetisation itself can be neglected and only the magnetisation gradients across the domain walls are relevant. The dielectric tensor is then simplified and we get
\begin{eqnarray}
\label{eqn-GradientTensor-simplified}
\vec{D} = P_\mathrm{gr}\left( \begin{array}{ccc}
	-\frac{\partial m_y}{\partial x} &\quad \frac{\partial m_x}{\partial x} \vspace{0.05in} \\
	-\frac{\partial m_y}{\partial y} &\quad \frac{\partial m_x}{\partial y} 
\end{array} \right) \vec{E}_{\mathrm{in}}\, .
\end{eqnarray}

This dielectric expression
 can now be applied to verify the domain boundary contrast in Fig.\,\ref{Fig-Voigteffect}. 
Let us take the marked walls in Fig.\,\ref{Fig-Voigteffect}(c) as an example. Here the light is polarised at $+45^\circ$, so the $x$- and $y$ components of the electrical field vector $\vec{E}_{\mathrm{in}}$ are equal. 
For domain wall (1) only the gradient $\frac{\partial m_x}{\partial y}$ is non-zero and negative. It thus causes a negative $D_y$ component that opposes the normally reflected $E_y$ component, leading to an amplitude decrease and consequently a dark domain boundary contrast. 
For domain wall (2) it is opposite --- here again only the gradient $\frac{\partial m_x}{\partial y}$ is non-zero but now positive. It thus leads to an increase of the $E_y$ component and a white boundary contrast. 
For boundary (3) we get $\frac{\partial m_x}{\partial x}, \frac{\partial m_x}{\partial y} > 0$ and $\frac{\partial m_y}{\partial y}, \frac{\partial m_y}{\partial x} < 0$. Consequently, all components of the dielectric tensor become positive so that both components of the reflected electrical field vector experience an amplitude increase, leading to a white boundary contrast. 
In case of wall (4) it is opposite: Here we find $\frac{\partial m_x}{\partial x}, \frac{\partial m_y}{\partial x} < 0$ and $\frac{\partial m_y}{\partial y}, \frac{\partial m_x}{\partial y} > 0$. The four components of the tensor therefore change sign alternately so that neither of the components of the $\vec{D}$ vector adds to the normally reflected field vector. This domain wall will therefore not show up with a contrast. Based on Eq.\,(\ref{eqn-GradientTensor-simplified}) the observed contrasts of all domain boundaries in Fig.\,\ref{Fig-Voigteffect} can be verified.


\section{Kerr Microscopy with separated paths}
\label{Overview Microscopy}

To check for eventualities, we have tried to verify our findings by looking for the same effects and contrasts in our microscope with separated illumination and reflection paths [Fig.\,\ref{Fig-Microscope}(d)], again by omitting the analyser. The results are summarised in Fig.\,\ref{Fig-Overview}:
\vspace{0.01in}
\begin{itemize}
\item 
The sensitivity curves for pure transverse Kerr microscopy could be verified, compare Fig.\,\ref{Fig-Transverse}(e) and Fig.\,\ref{Fig-Overview}(a). Also the domain contrast confirms pure transverse sensitivity --- compare Figs.\,\ref{Fig-Transverse}(a), (b) and Fig.\,\ref{Fig-Overview}(g). 
\item 
The quadratic MLD contrast, found for oblique incidence of $s$-polarised light, could be verified --- compare Fig.\,\ref{Fig-Transverse}(e) and Fig.\,\ref{Fig-Overview}(b). 
\item 
The longitudinal dichroic contrast, observed for plane-polarised light at an angle of $45^\circ$ relative to the plane of incidence, could also be verified --- compare the images and curves in Fig.\,\ref{Fig-Oppeneer} with the curves in Fig.\,\ref{Fig-Overview}(c), (d) and the images in Fig.\,\ref{Fig-Overview}(h), (i).
The presumed elliptical light contribution, generated by the beam splitter in our regular microscope, does not seem to have a significant influence for this effect.
\item 
The phase shift symmetry by using circularly polarised light, indicating the presence of transverse sensitivity with superimposed (weaker) longitudinal sensitivity could be verified --- compare the curves in Fig.\,\ref{Fig-Circular-curve} with Fig.\,\ref{Fig-Overview}(e), (f) and the domain images in Fig.\,\ref{Fig-Circular-domains} with Fig.\,\ref{Fig-Overview}(j).
\item 
On perpendicularly magnetised films we could verify the presence and symmetry of the MCD-based Kerr contrast [Fig.\,\ref{Fig-Overview}(k), (l)]. 
\item 
{The existence of a \emph{polar} dichroic contrast, expected for plane-polarised light at an angle of $\pm45^\circ$ relative to the incidence plane, could be verified on the garnet film, i.e. in Faraday geometry. As the domain width is below resolution, we have measured the hysteresis curve on this specimen in a magnetic field perpendicular to the film plane for demonstration [Fig.\,\ref{Fig-Overview}(m)]. For a polariser setting of zero degrees, no MO signal was found as expected.
However, we did not find any polar contrast for metallic films with perpendicular anisotropy, i.e. for Kerr geometry. The contrasts in Fig.\,\ref{Fig-CoFeB} (f, g) could thus not be reproduced. 
This indicates, as explained in Sect.\,\ref{Polar MCD-based Faraday microscopy}, that in this case the generation of elliptically polarised light due to the mirror in the beam splitter must be responsible for the contrasts in Fig.\,\ref{Fig-CoFeB} and not the
45$^{\circ}$-dichroic contrast. It seems that the polar 45$^{\circ}$-dichroic \emph{Kerr} contrast is not strong enough to be detected in contrast to the polar dichroic \emph{Faraday} contrast.}
\end{itemize}

\begin{figure}
\center
\includegraphics[width=1\linewidth,clip=true]{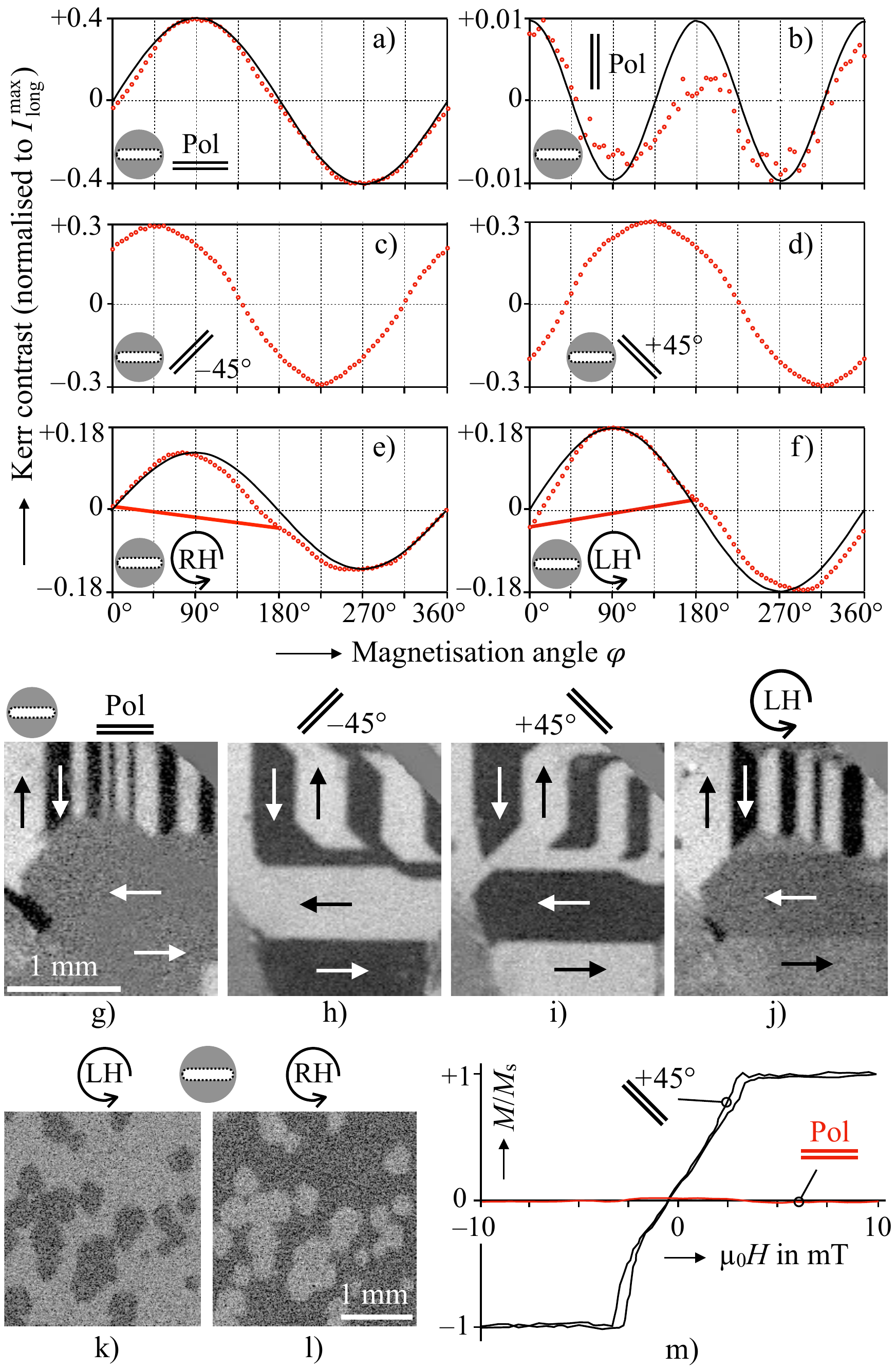}
\caption{
Measurements and low-resolution domain observations in the microscope with separated illumination and reflection paths. (\textbf{a}) - (\textbf{f}) Sensitivity curves on the FeSi specimen, measured in red light under the indicated conditions. (\textbf{g}) - (\textbf{j}) exemplary domain images, revealing the characteristic contrast features that are expected from the curves. (\textbf{k}), (\textbf{l}) Nucleated domains in the Pt/Co/Pt film with perpendicular anisotropy, observed in white, circularly {(better elliptically)} polarised light of opposite helicities. 
 {(\textbf{m}) Hysteresis curves of the garnet film in perpendicular field, measured with red, plane-polarised light at the indicated polariser settings.
For circular polarisation the 550\,nm quarter wave plate was applied.}
}
\label{Fig-Overview}
\end{figure}


\section{Further aspects}
\label{Further aspects}

 {Let us finally point out three further findings related to the previously described phenomena without going into much detail, i.e.\ leaving them open for future examinations.}

\subsection{{Influence of light colour} }
\label{Influence of light colour}

In Fig.\,\ref{Fig-Colour} we have collected some representative sensitivity curves that were obtained on the FeSi sample at the indicated conditions. The curves were measured with blue, red and white light and the intensities are normalised to the maximum intensity in each case. Therefore the fact that the absolute amplitude of the Kerr signal depends on the wavelength, as mentioned several time throughout the paper, is not visible in the graphs. It becomes obvious, however, that (besides the signal amplitude) also the phase of the measured curve can be more or less colour dependent. We furthermore need to point out that in our experiments we have used a broad-band quarter wave plate to generate circularly polarised light. Depending on the wavelength, the light can therefore be more elliptically than circularly polarised. It is thus expected that the MCD signals can be enhanced by using monochromatic light with a wavelength that is adapted to the Kerr spectra of the material under investigation together with wavelength-specific circular polarisers.

\begin{figure}
\center
\includegraphics[width=1\linewidth,clip=true]{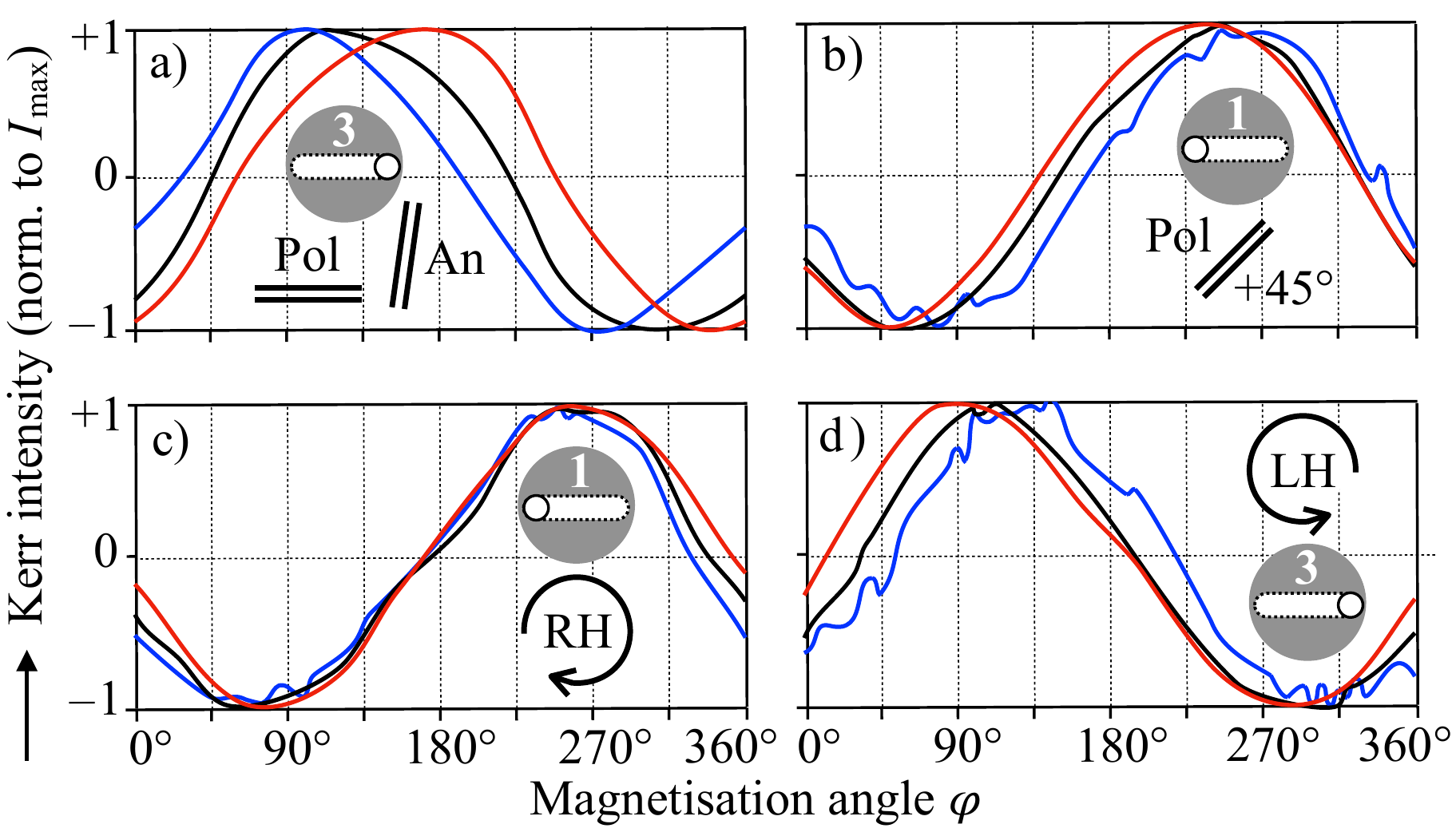}
\caption{
Selected sensitivity curves, measured on the FeSi sample under the indicated conditions by using red, blue and white light.
The maximum intensities have been equalised for all curves to emphasise the phase of the curves. In absolute numbers,
the contrasts in (a) are around 40\% of $C_\mathrm{max}^\mathrm{long}$ for white and blue light. In (b) to (d) they are around 12\% for red light an 7\% for blue and white light. {For circular polarisation the 550\,nm quarter wave plate was applied.}
}
\label{Fig-Colour}
\end{figure}


\subsection{{Imaging without analyser and polariser} }
\label{Imaging without analyser and polariser}

Figure\,\ref{Fig-Nopolarizer} demonstrates that domains in FeSi can even be seen without any polariser in the normal wide-field polarisation microscope. A closer inspection of the domain contrast reveals that the transverse Kerr effect dominates, which is understandable as this effect should exist even for unpolarised light. Nonetheless, a domain contrast was also seen on the perpendicularly magnetised garnet- and the Co/Pt/Co films in (our) normal wide-field microscope (not shown). The transverse contrast on the FeSi specimen could be verified in the microscope with separated paths [Fig.\,\ref{Fig-Microscope}(d)] after omitting also the polariser, thus confirming that the transverse Kerr effect does not require plane polarised light. However, in the microscope with separated paths no indication of domain contrast could be seen on the films with perpendicular anisotropy. The occurrence of polar contrast in our normal microscope may thus be attributed to the beam splitter [Fig.\,\ref{Fig-Microscope}(c)] which seems to be responsible for a net elliptical component of the incident light that causes a MCD-based polar Kerr effect. Furthermore it cannot be excluded that also the lenses in the microscope along the illumination path can generate elliptically polarised light, possibly caused by mechanical stress in the glasses. 

\begin{figure}
\center
\includegraphics[width=1\linewidth,clip=true]{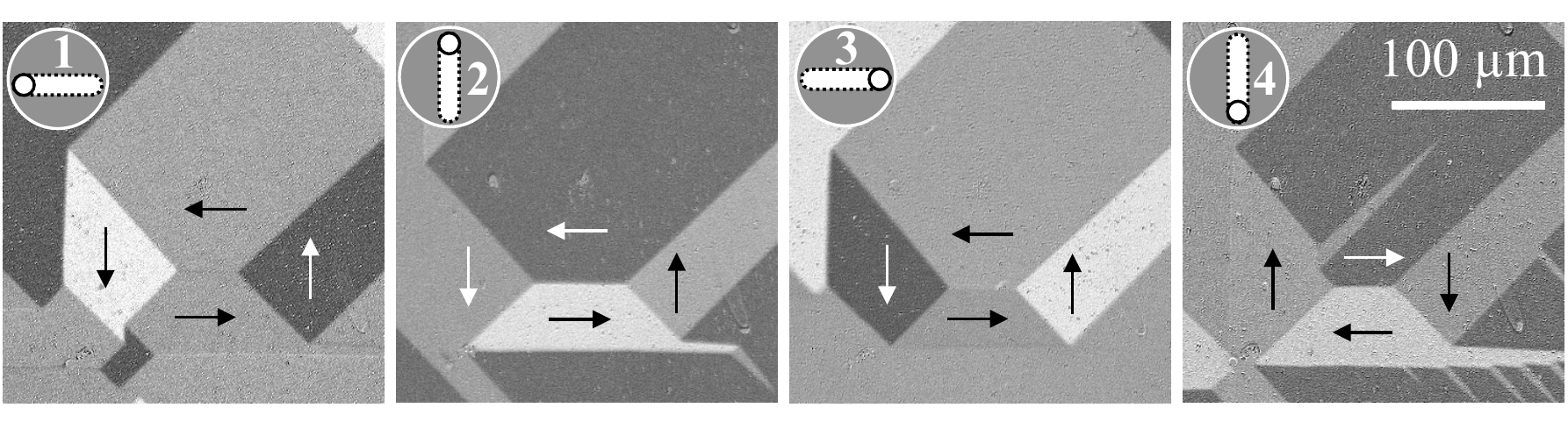}
\caption{
Domain images on the FeSi sheet, observed by using red light in the normal Kerr microscope but \textit{without analyser and polariser}.
The dependence of the contrast on the plane of incidence clearly indicates a dominance of the transverse Kerr effect.}
\label{Fig-Nopolarizer}
\end{figure}


\subsection{{Diffraction effects} }
\label{Diffraction effects}

{In Fig.\,\ref{Fig-MOIF-linear}\,(a) we have seen that the domain walls in our garnet film show up as a dark line contrast when imaged at a polariser setting of zero degrees in the absence of the analyser. In Fig.\,\ref{Fig-MOIF-Phasecontrast} (a) this contrast is shown again on a similar domain pattern, but now enhanced by background subtraction. By such enhancement, a black (though lower) line contrast is also seen for a polariser setting of 90$^{\circ}$ (not shown). The line contrast cannot be caused by the wall magnetisation itself as the Bloch-wall width in such garnet films is in the 10\,nm range, which is well below the resolution of 640\,nm for the given objective lens and wavelength.
Also the magneto-optical gradient effect cannot be responsible for this boundary contrast as any gradient contrast should change sign at every other wall.}

{Black domain boundary contrasts have already been observed by regular, rotation-based magneto-optical microscopy on in-plane magnetised iron films in the early 1960ies \cite{Boersch1964}, see Ref.\,[\onlinecite {Lambeck1977}] for a review. In Fig.\,\ref{Fig-MOIF-Phasecontrast}\,(b-d) the principle is documented for our garnet film by applying conventional, analyser-based Faraday imaging using plane-polarised light. As expected, the domain contrast is inverted by inverting the analyser opening direction [compare Fig.\,\ref{Fig-MOIF-linear}\,(b) and (c)]. This is due to polar Kerr vectors of opposite sign that are generated by the antiparallel domains and which are pointing perpendicular to the polariser axis thus causing clock- and counterclockwise rotations of the emerging wave (see Fig.\,\ref{Fig-Kerreffects}). If the analyser is crossed to the polariser [Fig.\,\ref{Fig-MOIF-Phasecontrast}\,(d)], the two Kerr amplitudes transmitted by the analyser have the same magnitude and the two domain phases will consequently reflect the light with equal intensities, thus not causing any domain contrast. However, the two opposite Kerr amplitudes are phase-shifted by 180$^{\circ}$. They therefore interfere destructively within a region around the walls that is determined by the lateral resolution, leading to dark lines at the domain boundaries.
As the line contrast is caused by phase shifts, it may be classified as \emph{phase contrast} \cite {Lambeck1977}. 
Note that this phase contrast does not depend on the wall magnetisation itself as long as the wall width is below resolution. 
The same is true for the gradient effect \cite{Schaefer1990} --- in both cases only the domain magnetisation on both sides of the domain walls is responsible for the domain boundary contrasts. If the analyser would be set parallel to the polariser, the Kerr amplitudes would be blocked so that the light would not experience any magnetisation-dependent amplitude or phase changes by passing the specimen.}

\begin{figure}
\center
\includegraphics[width=1\linewidth,clip=true]{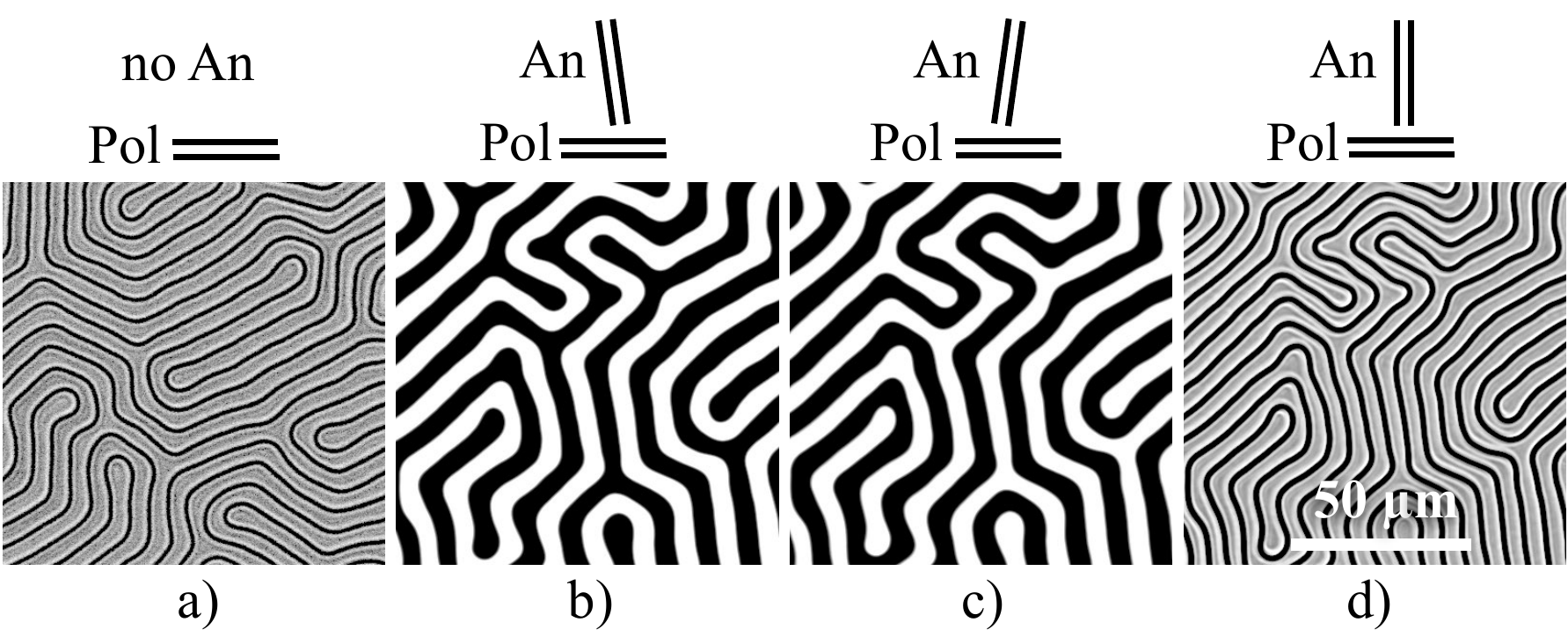}
\caption{{Magneto-optical phase- and domain contrast, imaged on the garnet film in plane-polarised, red light at perpendicular incidence: (\textbf{a}) Phase contrast without analyser, (\textbf{b, c}) conventional domain contrast with oppositely opened analyser, and (\textbf{d}) phase contrast with crossed analyser and polariser. 
Images (\textbf{a}) and (\textbf{d}) are difference images, for which a background image was subtracted to enhance the contrasts. For the ``non-magnetic" background image an AC magnetic saturation field perpendicular to the film plane was applied and the resulting, oscillating image intensity was averaged.}}
\label{Fig-MOIF-Phasecontrast}
\end{figure}

{
The domain boundary contrast in Fig.\,\ref{Fig-MOIF-Phasecontrast}\,(a) can be interpreted along the same line: 
We have seen that the prerequisites for phase contrast at domain boundaries is the presence of phase-shifted Kerr (or Faraday-) amplitudes and the absence of domain contrast. The latter was achieved by the crossed analyser in case of Fig.\,\ref{Fig-MOIF-Phasecontrast}\,(d). In the geometry of Fig.\,\ref{Fig-MOIF-Phasecontrast}\,(a) the Faraday amplitudes are the same as in the previous case and the absence of domain contrast is intrinsically given. The domain boundary contrast in (a) may consequently be interpreted as \emph{phase contrast}.
}

{
It should also be noted that the phase contrast depends on the illumination aperture \cite {Lambeck1977}: Fresnel diffraction fringes may show up for small apertures, while other contrast artefacts may be superimposed in case of oblique light incidence. In Ref.\,[\onlinecite{Schaefer1990}] is was shown that a kind of {schlieren} effect may occur if perpendicularly magnetised domains are observed at oblique incidence with small aperture and if the reflected beam passes close to an edge of the aperture. The mentioned phase jump then leads to differently deflected beams at neighbouring domain walls which are either cut off by the aperture or which are deflected into the aperture, thus resulting in an alternating bright and dark appearance of neighbouring domain boundaries that may be superimposed to the polar gradient effect. Prerequisite is again the suppression of domain contrast, e.g.\ by crossed polariser and analyser.
}

{Talking about diffraction effects at domain boundaries, let us finally mention \emph{dark-field microscopy}. In the previous discussions we have seen that the light, passing a domain structure, consists of a normal amplitude and a magnetisation-dependent Kerr- or Faraday amplitude that is polarised perpendicular to the incoming, plane polarised wave.  As the normal amplitude does not contain magnetic information, it needs to be suppressed or at least reduced. In conventional, bright-field Kerr- and Faraday microscopy this is achieved by an (almost) crossed analyser. The presence of a magneto-optical amplitude, however, is independent of the presence or absence of an analyser. Those components will as well be present if the normal component is suppressed by running an optical polarisation microscope in the dark-field mode 
In Refs.\,[\onlinecite{Lambeck1964}], [\onlinecite{Lambeck1977}] and [\onlinecite{Schaefer1990b}] it was shown that in dark-field microscopy the domain boundaries show up as \emph{bright lines} even in the absence of any polarising element. Adding polariser and analyser may help to separate those magnetic line contrasts from superimposed contrast emerging form non-magnetic, topographic features like scratches that show up under the same conditions. In Fig.\,\ref{Fig-MOIF-Darkfield} the dark-field contrast was reproduced for our garnet film without using any polarising element.
 }
 
\begin{figure}
\center
\includegraphics[width=1\linewidth,clip=true]{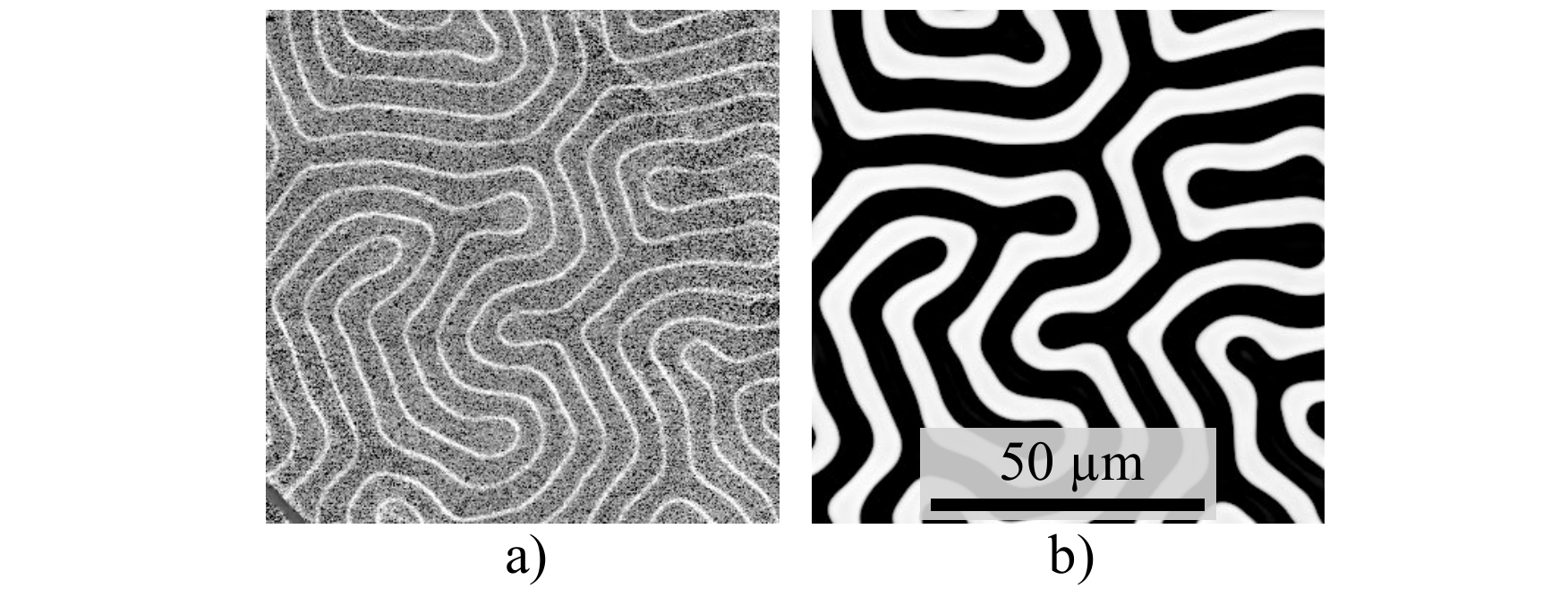}
\caption{{(\textbf{a}) Dark-field image of the garnet film, observed in white light by means of a Carl Zeiss dark-field reflector and a 20x HD DIC objective without polariser and analyser. Shown is a difference image to enhance the contrast. (\textbf{b}) Corresponding regular Faraday image, obtained in red light by means of the analyser.}}
\label{Fig-MOIF-Darkfield}
\end{figure}


\section{Conclusions}

Different to conventional magneto-optical Kerr and Faraday microscopy, we have demonstrated that it is also possible to image magnetic domains in an optical wide-field polarisation microscope without needing an analyser. {\red We have shown that} using plane-polarised light for illumination, the transverse Kerr effect can be employed on in-plane magnetised materials when the light has a $p$-component and the magnetisation has a component perpendicular to the incidence plane being maximum for $\vec{k} \perp \vec{m}$ with $\vec{k}$ the wavevector.
We have {\red furthermore} shown that the transverse Kerr effect is even active in the absence of the polariser thus not requiring plane-polarised light at all.
In circularly- or elliptically polarised light, materials with in- and out-of-plane magnetisation can be imaged when $(\vec{k} \cdot \vec{m}) \neq 0$.
Here the domain contrast is caused by magnetisation-dependent intensity modulations due to the Magnetic Circular Dichroism (MCD).
This works in both, transmission (Faraday) as well as reflection (Kerr) geometry. 
A further magnetic intensity contrast can be obtained for longitudinal in-plane magnetisations, using plane-polarised light at a nonzero angle $\theta$ {($\approx \pm 45^{\circ}$)} with respect to the incidence plane, also when $(\vec{k} \cdot \vec{m}) \neq 0$.
{\red The Magnetic Linear Dichroism (MLD), lastly, provides a weaker contrast mechanism than the MCD, but it could be applied for situations where the MCD is not effective, for example, the visualisation of antiferromagnetic domains.}  

{\red While our study draws attention to intensity-modulation effects for imaging, it deserves to be noted that} even in conventional Kerr microscopy one needs to consider that {\red such} effects
may be superimposed to the conventional contrasts. This is most obvious for the transverse Kerr effect, which may lead to phase-shifted sensitivity curves in case of the longitudinal Kerr microscopy with $s$-polarised light as demonstrated in Fig.\,\ref{Fig-Conventional}. But also the curve distortions, visible in that figure, may be related to superimposed quadratic effects or MCD effects being caused by elliptical light contributions due to the beam splitter or stressed optical components in the microscope. {\red Also the phase contrast at domain boundaries (Fig.\,\ref{Fig-MOIF-Phasecontrast}) may be superimposed to conventional domain images at insufficient analyser opening angles.} 

While the MCD effect, induced by magnetic fields at visible light frequencies, plays an established role in the investigation of molecular electronic structure and transitions by using spectroscopic methods \cite{Mason2007}, its potential for magnetic imaging has apparently been disregarded so far. MCD microscopy with visible light thus provides an interesting alternative to the more elaborate techniques that are based on XMCD 
\cite{vonKorff2014,Zayko2020}.
Provided that the amplitude modulations are large enough, analyser-free wide-field MO microscopy would also offer some advantages compared to conventional, analyser- and compensator-based microscopy: (i) The analyser and the compensator, necessary for contrast optimisation in a conventional MO microscope, can be omitted thus reducing the number of optical elements and with it the complexity of contrast adjustment.  (ii) In conventional MO microscopy one usually works at almost crossed polariser and analyser, i.e.\ the overall image intensity is rather low requiring light sources with a high luminous density, polarisers with high transmittance, and advanced cameras with high sensitivities to achieve domain images with good signal-to-noise ratios. In analyser-free MO microscopy the requirements for light source, polarisers and cameras are much more relaxed as the image brightness posses no problem. (iii) The maximum domain contrasts in the analyser-free modes are reduced by a factor of approx. ten compared to the best contrasts obtained by conventional longitudinal and polar Kerr microscopy. {Using monochromatic light and wave-length specific circular polarisers helps to improve the signal in case of the MCD effect.} 
{\red For quantitative Kerr microscopy, which relies on well-defined sensitivity curves to calibrate the domain contrast, the pure transverse Kerr effect (see Fig.\,\ref{Fig-Transverse}) seems to be the best choice.}
{\red Overall, we find that,} enhanced by background subtraction, the achievable contrast for both, in-plane as well as perpendicularly magnetised media does not suffer significantly compared to conventional domain contrasts. 
%
%

{\red Summarising, our study highlights possible approaches for performing magnetic domain imaging in intensity-based, wide-field magneto-optical microscopy.  Implementing these approaches can make magnetic domain imaging widely available in simplified microscopy setups.}


%
%

%

\begin{acknowledgments}
R.S.\ acknowledges stimulating discussions with Vladislav Demidov (M\"{u}nster).
I.S.\ is grateful to the DFG (German Research Foundation) for supporting this work through project SO 1623/2-1 and P.M.O.\ acknowledges support from the Swedish Research Council (VR). 
A.O.\ and  A.S.\ thank the Russian Ministry of Science and Higher Education for state support of scientific research conducted under the supervision of leading scientists in Russian institutions of higher education, scientific foundations and state research centres (Project proposal No.\ 2020-220-08-4899), for the state task (0657-2020-0013) and the Act 211 of the Government of the Russian Federation (02.A03.21.0011). We thank  Prof.\ Young Keun Kim (Seoul) and his group for providing the Ta/CoFeB/MgO/Ta film.
Technical support by Stefan Pofahl (IFW) is acknowledged. 
Special thanks to Heiner Eschrich {\red and Michael Z\"{o}lffel} (Carl Zeiss AG, Jena) for valuable advise on optical microscopy. 
The authors thank Michal Urbanek (Brno) for providing the Pt/Co/Pt film.
\end{acknowledgments}

The data that support the findings of this study are available from the corresponding author upon reasonable request.


%

\end{document}